\documentclass{iopjournal}

\usepackage{amsmath}
\usepackage{amssymb}
\usepackage{braket}
\usepackage{physics}
\usepackage{booktabs}
\usepackage[expansion=false]{microtype}
\usepackage[shortcuts]{glossaries}
\usepackage{subcaption}
\usepackage{enumitem}
\usepackage{comment}
\usepackage{sidecap}
\sidecaptionvpos{figure}{b}

\hypersetup{colorlinks=true, allcolors=black}

\makeglossaries

\newacronym{scm20}{SCM-20}{20-node supply-chain model}

\newacronym{qae}{QAE}{Quantum Amplitude Estimation}
\newacronym{mlae}{MLAE}{Maximum Likelihood Amplitude Estimation}
\newacronym{iqae}{IQAE}{Iterative Quantum Amplitude Estimation}
\newacronym{qpe}{QPE}{Quantum Phase Estimation}
\newacronym{qpu}{QPU}{Quantum Processing Unit}
\newacronym{nisq}{NISQ}{Noisy Intermediate-Scale Quantum}

\newacronym{bp}{BP}{Loopy Belief Propagation}
\newacronym{bpacl}{BPCL}{Belief-Propagation-Assisted Chow--Liu}
\newacronym{mcmc}{MCMC}{Markov Chain Monte Carlo}
\newacronym{cvar}{CVaR}{Conditional Value at Risk}

\newacronym{naivemc}{IMA}{Independent-Marginal Approximation}
\newacronym{fixedis}{FIS}{Fixed Importance Sampling}
\newacronym{gibbsmc}{GMC}{Gibbs Markov Chain Monte Carlo}
\newacronym{ice}{iCE}{iterative Cross-Entropy}

\newcommand{\KL}[2]{D_\mathrm{KL}(#1 \| #2)}

\begin{document}

\articletype{Paper}

\title{Quantum Rare-Event Estimation for Ising Graphical Models with Belief-Propagation State Preparation}

\author{Mario Hern\'andez Vera$^{1}$, Narges Jamialahmadi$^{2}$, Burak Mete$^{1}$,  Susanne Schneider$^{2}$ and Marco De Pascale$^{1}$}

\affil{$^1$Munich Quantum Valley MQV gGmbH, Walther-von-Dyck-Straße 6, 85748 Garching bei München, Germany}

\affil{$^2$SAP Labs Munich (MUE03), Friedrich-Ludwig-Bauer-Stra{\ss}e 5, 85748 Garching bei München, Germany}

\email{mario.hernandezvera@munich-quantum-valley.de}


\begin{abstract}

Quantum amplitude estimation can reduce the sampling cost of rare-event probability
estimation, but applying it to correlated Ising graphical models is limited by the
difficulty of preparing the target distribution and building a practical event oracle.
This work explores two approximate strategies for mitigating these challenges.
We introduce a sample-free state-preparation method combining loopy belief
propagation with the Chow--Liu algorithm. The resulting tree approximation is
compiled into a quantum circuit with linear gate count and depth, and its accuracy
is evaluated across graph families spanning different topologies, coupling
strengths, and coupling signs. We also construct a structural oracle that evaluates
threshold rules with reversible Boolean gates. Using a twenty-node supply-chain
disruption model as a case study, we compare maximum likelihood amplitude
estimation against four classical Monte Carlo baselines. Under the fixed-depth
schedule used throughout this work, the quantum estimator has the same asymptotic
error scaling as the classical methods but achieves lower estimation error by a
constant factor. This reduction narrows when amplitude-encoding queries replace
raw shots as the resource metric. We separate statistical error from the
deterministic errors caused by approximate state preparation and oracle
construction, and identify the requirements for achieving an improvement beyond
a constant factor.

\end{abstract}

\section{Introduction}

Supply chains are exposed to rare but financially severe disruptions,
including supplier failures, logistics bottlenecks, demand shocks, and
geopolitical events~\cite{Katsaliaki:2022, Lukasz:2025}. A natural risk
metric is the tail probability $P(C>T)$ that the total disruption cost $C$
exceeds a critical threshold $T$. In realistic industrial models, the
number of possible disruption states grows exponentially with the number
of uncertain factors. Exhaustive evaluation therefore becomes infeasible
even for moderately sized systems.
The most decision-relevant outcomes often lie in the tail of the cost
distribution. Their probabilities can be small and costly to estimate
accurately. Standard classical Monte Carlo (MC) estimators have variance
$O(N^{-1})$ and standard error $O(N^{-1/2})$, where $N$ is the number of
samples~\cite{Woerner:2019, Herman:2023}. Variance-reduction methods such
as importance sampling and iterative cross-entropy can improve the
constant factor, but they do not change this asymptotic convergence rate.

\ac{qae}~\cite{Brassard:2002} can reduce the number of oracle queries
required to estimate a probability. In its ideal form, the mean squared
error scales as $O(N_\mathrm{query}^{-2})$, where $N_\mathrm{query}$ is the
number of oracle queries, rather than the classical $O(N^{-1})$. The
corresponding standard error scales as $O(N_\mathrm{query}^{-1})$ instead
of the classical $O(N^{-1/2})$~\cite{Montanaro:2015}. Applications have
been studied in
financial risk analysis~\cite{Woerner:2019}, derivative
pricing~\cite{Rebentrost:2018, Stamatopoulos:2020, Chakrabarti:2021},
insurance, and network risk models~\cite{Braun:2023}.
The canonical form of \ac{qae} relies on \ac{qpe}. It requires deep
controlled circuits and additional ancilla qubits, which makes it
impractical for near-term hardware. \ac{qpe}-free methods such as
\ac{mlae}~\cite{Suzuki:2020} and
\ac{iqae}~\cite{Grinko:2021} avoid these requirements. They can recover
the ideal asymptotic rate when the maximum Grover depth grows with the
available query budget.

Applying \ac{qae} to this discrete, correlated supply-chain risk model also
requires two practical circuit components.
The first is a state-preparation unitary
$\mathcal{R}$. It must prepare a quantum state whose amplitudes encode the
Boltzmann probabilities induced by the local risk factors and pairwise
dependencies of the model. Constructing this state efficiently is
nontrivial for large correlated networks. Exact preparation of an
arbitrary $n$-qubit state requires exponentially many
gates~\cite{Plesch:2011, Zoufal:2019, Chakrabarti:2021}, which can remove
the potential advantage of the estimation procedure.

Existing state-preparation methods do not directly resolve this problem.
Quantum Generative Adversarial Networks (qGAN) require offline training and have
mainly been demonstrated for small scalar distributions~\cite{Zoufal:2019},
while quantum circuit Markov random fields still rely on gradient-based
training against samples or a target distribution~\cite{Bako:2026}.
Matrix Product State methods provide efficient deterministic circuits for
smooth continuous distributions, but do not directly address the structured
Boltzmann distributions considered here~\cite{Rattew:2021, Iaconis:2024}.
The Grover-Rudolph algorithm also
requires efficient classical evaluation of conditional marginals, which
for general Ising Boltzmann distributions involves difficult partial
partition functions~\cite{Grover:2002}. Estimating these quantities with
classical Monte Carlo can remove the speedup of the combined quantum
pipeline~\cite{Herbert:2021}. Bounded-treewidth Ising graphs (treewidth
is a graph invariant measuring how tree-like a graph is) are an
important special case because junction-tree methods can compute their
marginals in time polynomial in the number of variables and exponential
in the treewidth~\cite{KollerFriedman:2009}. These marginals could in
principle be converted into conditional rotation angles using ideas from
bounded-in-degree Bayesian networks~\cite{Low:2014, Borujeni:2021}, but
the cost grows rapidly with treewidth and no clear construction has yet
been developed for general Ising or Markov random field distributions.

The second required component is the event oracle
$\mathcal{F}_T$. It must mark every state $x$ for which $C(x)>T$.
Comparator and reversible arithmetic circuits have been proposed for
Value-at-Risk estimation~\cite{Woerner:2019}, path-dependent
derivatives~\cite{Stamatopoulos:2020, Chakrabarti:2021,
CarreraVazquez:2021}, network cascade costs~\cite{Braun:2023}, and polynomial binary
objectives~\cite{Gilliam:2021}. These constructions assume that the cost
function can be represented by a tractable arithmetic expression, a
known comparator, or a polynomial objective.
The supply-chain cost model considered here also contains discrete
Boolean activation rules. A direct diagonal implementation would require
the cost of all $2^n$ configurations to be computed classically before
the quantum circuit is constructed. Such an oracle is useful as a
simulation reference, but it is not a scalable \ac{qpu} implementation.
Recent work on quantum rare-event estimation focuses on extreme-event
discovery rather than fixed-oracle tail-probability
estimation~\cite{Guo:2026}, and does not address this circuit-construction
problem.

This work explores two approximate strategies for the main
implementation challenges of the quantum pipeline. We introduce the \ac{bpacl}
method, that uses classical \ac{bp} to estimate the relevant marginals of
the Ising model~\cite{Yedidia:2003, Murphy:1999}. These marginals
define a Chow-Liu tree approximation to the Boltzmann
distribution~\cite{ChowLiu:1968}, which is then compiled into a
quantum state-preparation circuit with linear gate count and depth. We
test this accuracy across a range of graph topologies of increasing
complexity, not only the specific benchmark used for the rest of the
paper. A structural oracle evaluates the relevant supply-chain
threshold rules directly on the quantum register. Its threshold is derived from the model structure,
so the construction avoids enumerating the costs of all possible
configurations. The analysis separates approximation errors from sampling
errors and evaluates computational cost from several complementary
perspectives. The numerical study compares \ac{mlae} with \ac{naivemc},
\ac{gibbsmc}, \ac{fixedis}, and \ac{ice} at several budgets and for two
oracle constructions. The results are reported in terms of quantum shots,
amplitude-encoding queries, and gate-level circuit resources, providing a
more balanced assessment than any single cost metric.
This work also uses \ac{mlae} with the fixed depth schedule
$k\in\{0,1,2,3\}$. This choice limits circuit depth but does not produce
the ideal quadratic asymptotic scaling. As shown in
Sec.~\ref{sec:results_schedule_scaling}, the resulting standard error
scales as $O(N^{-1/2})$. 
The Grover depths improve only the constant factor, while the asymptotic
exponent remains unchanged. Accordingly, the numerical results quantify
statistical efficiency under a bounded-depth schedule and do not demonstrate
the ideal \ac{qae} scaling.

The paper is organised as follows.
Section~\ref{sec:model} introduces the supply-chain Ising risk model and
the tail-probability metric.
Section~\ref{sec:classical} describes the four classical Monte Carlo
baselines.
Section~\ref{sec:qae} presents the general \ac{qae} pipeline.
Sections~\ref{sec:state_prep} and~\ref{sec:oracle_design} describe the
\ac{bpacl} state-preparation method and the two oracle constructions.
Section~\ref{sec:budget} defines the resource-accounting framework.
Section~\ref{sec:results} presents the numerical results.
Section~\ref{sec:discussion} discusses the scope of the potential quantum
utility and the scaling prospects.
Derivations of the \ac{bp} equations and the Chow-Liu circuit are provided
in Appendices~\ref{app:bp} and~\ref{app:chow_liu}.

\section{Supply Chain Risk Model}
\label{sec:model}

Modern supply networks are exposed to disruption risks that propagate across
tiers: a geopolitical shock, an energy crisis, or a port closure can raise
the failure probabilities of multiple downstream entities simultaneously.
Capturing such correlated, cascading failures requires a joint probability
model over the entire network state, not merely a collection of independent
marginal probabilities.
We meet this requirement by formulating the supply-chain disruption problem as
a binary Ising model, which provides a compact joint distribution with
calibrated marginal failure probabilities and pairwise dependencies and is
natively compatible with quantum sampling hardware.
The remainder of this section defines the probabilistic model
(Section~\ref{sec:ising}), the economic cost function and tail risk metrics
(Section~\ref{sec:cost}), and the specific calibrated benchmark instance used
throughout the paper (Section~\ref{sec:benchmark}).

\subsection{Ising Formulation of Disruption Risk}
\label{sec:ising}

Each supply-chain entity $i \in \{1,\ldots,n\}$ is represented by a binary
variable $x_i \in \{0,1\}$, where $x_i = 1$ indicates disruption and
$x_i = 0$ normal operation.
A complete network disruption scenario is a vector
$x \in \{0,1\}^n$, and we model its probability as a Boltzmann
distribution over the binary hypercube,
\begin{equation}
  P(x) = \frac{1}{Z}\exp\!\Bigl(
    \sum_{i} h_i\,x_i
    + \sum_{(i,j)\in E} J_{ij}\,x_i x_j
  \Bigr),
  \label{eq:boltzmann}
\end{equation}
where $Z$ is the partition function that normalises the distribution and
$E$ is the set of edges connecting statistically dependent node pairs.

The \emph{local field} $h_i$ controls the marginal failure probability of
node~$i$.
In the absence of couplings, $P(x_i=1) = \sigma(h_i)$, the logistic sigmoid,
so $h_i = \ln[p_i/(1-p_i)]$ is the log-odds of the marginal failure
probability~$p_i$.
Since all target probabilities lie between 2\% and 6\%, all fields are large
and negative: a more negative field means a lower baseline failure probability.

The \emph{coupling} $J_{ij}$ encodes statistical dependence between nodes $i$
and~$j$.
Its effect is transparent from the conditional log-odds,
\begin{equation}
  \ln\frac{P(x_i=1\mid x_{\partial i})}
         {P(x_i=0\mid x_{\partial i})}
  = h_i + \sum_{j\in\partial i} J_{ij}\,x_j,
  \label{eq:conditional}
\end{equation}
which shows that each disrupted neighbour $x_j=1$ shifts the log-odds of $x_i$
by~$J_{ij}$.
A positive coupling therefore raises the conditional failure probability of $i$
whenever a connected neighbour has already failed, directly encoding cascade risk.

Setting $h_i = \ln[p_i^{\mathrm{target}}/(1-p_i^{\mathrm{target}})]$ is exact
only when all couplings vanish.
Once the coupling graph is added, the marginals shift away from their targets.
We correct this with an iterative exact-enumeration scheme: at each step the
exact marginal $p_i^{(k)}$ is computed by full enumeration over all $2^n$
states, and the field is updated as
\begin{equation}
  h_i^{(k+1)} = h_i^{(k)} + \ln\!\Bigl(\frac{p_i^{\mathrm{target}}}{p_i^{(k)}}\Bigr).
  \label{eq:calib}
\end{equation}
If the model over-estimates the marginal of node~$i$, the logarithm is
negative and pushes $h_i$ downward. If the marginal is under-estimated,
the update instead pushes $h_i$ upward.

The log-partition function is strictly convex in the fields because the
sufficient statistics $x_1,\ldots,x_n$ are linearly independent.
Consequently, its gradient map $h\mapsto p(h)$ is injective. If a field
vector matching all $n$ target marginals exists, it is therefore unique.
This convexity argument does not by itself establish that the iteration
in Eq.~\eqref{eq:calib} converges to that vector.
In practice the procedure converges in six iterations to a maximum marginal
error below $10^{-8}$ for this 20-variable model, which is what we rely
on here.
This calibration step is used only to construct the specific benchmark
instance studied in this paper.
It is a one-time offline step, separate from \ac{bpacl} itself.
\ac{bpacl} only needs the resulting fields and couplings as input, and
requires no enumeration or sampling at runtime (Sec.~\ref{sec:state_prep}).

\subsection{Cost Function and Tail Risk Metrics}
\label{sec:cost}

Disruption costs in real supply chains are superadditive: simultaneous failures
remove redundancy options and can trigger non-linear operational escalation.
We capture this with a three-layer cost function,
\begin{equation}
  C(x)
  = \sum_{i} c_i\,x_i
  + \sum_{(i,j)\in\mathcal{P}} \pi_{ij}\,x_i x_j
  + \sum_{r=1}^{2} \Delta_r\,\mathbf{1}[\mathrm{rule}_r(x)].
  \label{eq:cost}
\end{equation}
The first term accumulates the direct cost $c_i$ of each individually
disrupted node.
The second term adds a superadditive penalty $\pi_{ij}$ whenever two
operationally interdependent nodes fail simultaneously, reflecting the
compounded loss from losing recovery options.
The third term fires a fixed surcharge $\Delta_r$ whenever the scenario
satisfies a structural predicate $\mathrm{rule}_r$ --- for instance when three or more
suppliers in the same tier fail concurrently, exhausting all contingency
capacity.
Together these layers produce a multimodal, non-Gaussian cost distribution with
sharp jumps in the tail, which is what makes rare-event estimation hard for
standard Monte Carlo.

The primary risk metric is the tail exceedance probability,
\begin{equation}
  p_{\mathrm{tail}}
  = P\bigl(C(x)>T\bigr)
  = \sum_{x} P(x)\,\mathbf{1}[C(x)>T].
  \label{eq:ptail}
\end{equation}
We use two complementary thresholds.
$T_\mathrm{ref}$ is the exact 97.5th percentile of the cost distribution
under the calibrated model, making $p_{\mathrm{tail}}\approx 2.4\%$ --- a
genuinely rare and therefore challenging estimation target.
$T_\mathrm{struct}$ is chosen immediately below a lower bound on the minimum
cost of a rule-firing state;
it depends only on the cost structure and marks the natural boundary between
a degraded but manageable state and a catastrophic operational failure.

\subsection{Benchmark Model Instance}
\label{sec:benchmark}

We use a calibrated $n=20$ binary Ising model as our benchmark. Its
$2^{20}=1\,048\,576$ states can be enumerated exactly, providing bit-exact
ground truth for all risk metrics without any Monte Carlo error in the
reference values.
The 20 nodes span five operational layers of a global manufacturing supply
chain: five Tier-1 direct suppliers, five Tier-2 sub-suppliers, three
continental manufacturing plants, five logistics assets (ports, air cargo,
ocean freight, rail), and two macro-level shock drivers representing an Asia
geopolitical shock and an EU energy crisis.

The coupling graph has $n=20$ nodes and $|E|=20$ edges. Rather than being a
single connected near-tree, it decomposes into four disconnected components:
a fifteen-node core sub-network with 18 edges and cyclomatic number 4
(four independent cycles), two
disjoint two-node pairs, and one isolated node. Despite the four independent
cycles, the core component has a series--parallel structure: it can be
constructed by repeatedly combining smaller graphs in series or parallel.
This structure gives the graph treewidth exactly 2.
The two macro-shock nodes act as high-degree hubs with the strongest couplings
and are the primary source of systemic correlation in the model.
The cost function includes eight pairwise penalties and two threshold rules,
producing a multimodal tail distribution that is non-trivial to sample
efficiently.

\section{Methods}
\label{sec:methods}

\subsection{Classical Monte Carlo Methods}
\label{sec:classical}

We benchmark four classical estimators for the tail probability
$p_{\mathrm{tail}} = P(C(x) > T)$ defined in
Sec.~\ref{sec:cost}: an independent-marginal approximation (\ac{naivemc}), Gibbs
sampling (\ac{gibbsmc}), fixed-shift Importance Sampling (\ac{fixedis}),
and iterative Cross-Entropy (\ac{ice}).
The four methods differ in how they draw samples from---or relative
to---the Boltzmann distribution~\eqref{eq:boltzmann}, and they span
the full spectrum from the simplest possible estimator to the most
adaptive one~\cite{Rubino:2009,Bucklew:2004,Juneja:2006}.
All four are evaluated on the \ac{scm20} benchmark described in
Sec.~\ref{sec:benchmark}, against the exact ground truth obtained
by full enumeration.

\subsubsection{Independent-Marginal Approximation}
\label{sec:naivemc}

The simplest estimator draws $N$ independent samples from the
\emph{product of marginals},
\begin{equation}
  Q_0(x) = \prod_{i=1}^{n} \bigl(p_i^{\mathrm{target}}\bigr)^{x_i}
                    \bigl(1-p_i^{\mathrm{target}}\bigr)^{1-x_i},
  \label{eq:naivemc_proposal}
\end{equation}
treating each node as independent with its calibrated marginal failure
probability $p_i^{\mathrm{target}}$ (Sec.~\ref{sec:ising}) --- the true
single-node marginal of the coupled model $P$, not the uncoupled
logistic reading $\sigma(h_i)$ of the calibrated field, which differs
from $p_i^{\mathrm{target}}$ once couplings are present.
For each sample $x^{(k)}$ the cost $C(x^{(k)})$ is
evaluated and the tail probability is estimated as the empirical
exceedance frequency,
\begin{equation}
  \hat{p}_{\mathrm{tail}}^{\mathrm{IMA}}
  = \frac{1}{N} \sum_{k=1}^{N} \mathbf{1}\bigl[C(x^{(k)}) > T\bigr].
  \label{eq:naivemc_est}
\end{equation}
Since $x^{(k)}$ is drawn from $Q_0$ rather than the true distribution
$P$, the indicator is Bernoulli with mean
$q_{\mathrm{tail}} = Q_0(C>T)$, not $p_{\mathrm{tail}} = P(C>T)$.
The standard error of the estimator therefore scales as
\begin{equation}
  \mathrm{SE} = \sqrt{\frac{q_{\mathrm{tail}}(1-q_{\mathrm{tail}})}{N}}
              \approx \frac{\sqrt{q_{\mathrm{tail}}}}{\sqrt{N}},
  \label{eq:naivemc_se}
\end{equation}
showing the familiar $1/\sqrt{N}$ convergence rate around
$q_{\mathrm{tail}}$. Since $q_{\mathrm{tail}}$ is close to
$p_{\mathrm{tail}} \approx 2.4\%$ (see below), even $N=1\,000$ samples
yield a relative standard error of roughly 20\%, making the estimator
highly variable at moderate budgets, on top of the permanent bias
discussed next.

Beyond variance, \ac{naivemc} carries a \emph{permanent bias}, since it samples
from $Q_0$ rather than the true Boltzmann distribution $P$.  Because
$Q_0$ uses the exact single-node marginals of $P$ but drops all
pairwise dependence, this bias arises solely from ignoring the
couplings $J_{ij}$, not from any error in the individual marginals.
Because the couplings are positive, the true distribution assigns more
probability mass to correlated failure configurations than $Q_0$ does,
so the estimator systematically underestimates $p_{\mathrm{tail}}$.

\subsubsection{Gibbs Sampling}
\label{sec:gibbsmc}

Gibbs sampling constructs a Markov chain that converges to the exact
Boltzmann distribution $P(x)$~\cite{Geman:1984}, eliminating the coupling bias
of \ac{naivemc}.
At each step, a single variable $x_i$ is selected (in fixed or random
order) and resampled from its full conditional distribution given all
other variables.
From Eq.~\eqref{eq:conditional}, this conditional is a Bernoulli
with success probability
\begin{equation}
  P(x_i = 1 \mid x_{\partial i})
  = \sigma\!\Bigl(h_i + \sum_{j \in \partial i} J_{ij}\,x_j\Bigr),
  \label{eq:gibbs_conditional}
\end{equation}
which is cheap to evaluate because each node has at most a small number
of neighbours in the sparse coupling graph.
After a burn-in period of $B$ steps that allows the chain to
reach stationarity, the estimator is
\begin{equation}
  \hat{p}_{\mathrm{tail}}^{\mathrm{GMC}}
  = \frac{1}{N} \sum_{k=1}^{N} \mathbf{1}\bigl[C(x^{(k)}) > T\bigr],
  \label{eq:gibbs_est}
\end{equation}
where $x^{(1)}, \ldots, x^{(N)}$ are the post-burn-in
states of the chain.

Because consecutive samples are correlated, the nominal count $N$
overstates the true information content of the run.
The effective sample size is
\begin{equation}
  N_{\mathrm{eff}} = \frac{N}{2\tau},
  \label{eq:neff}
\end{equation}
where $\tau = \tfrac12 + \sum_{t\ge1}\rho_t$ is the \emph{integrated
autocorrelation time} of the indicator $\mathbf{1}[C(x) > T]$, in
Sokal's windowed-estimator
convention~\cite{Sokal:1989,Geyer:1992,Vehtari:2021}.
A large $\tau$ means the chain mixes slowly and consecutive samples
carry nearly the same information, so the actual statistical efficiency
is much lower than the raw sample count suggests.

\subsubsection{Importance Sampling}
\label{sec:fixedis}

Importance Sampling (IS) replaces the nominal distribution $P$ with a
proposal distribution $Q_\mu$ that is tilted toward high-cost
configurations, reducing the variance of the tail estimator~\cite{Juneja:2006,Bucklew:2004}.
We use \emph{exponential tilting} with an \emph{independent} proposal:
each node is sampled independently with its log-odds shifted by $\mu > 0$,
\begin{equation}
  Q_\mu(x)
  = \prod_{i=1}^{n}
    \sigma(h_i+\mu)^{x_i}
    \bigl[1-\sigma(h_i+\mu)\bigr]^{1-x_i},
  \label{eq:is_proposal}
\end{equation}
which uniformly inflates the marginal failure probability of every node
and thereby oversamples the costly tail region.
The key design choice is that $Q_\mu$ ignores all couplings $J_{ij}$.
This makes sampling trivial ($O(n)$ per draw, fully independent), whereas
drawing from a full Ising proposal with shifted fields would require
MCMC and reintroduce the mixing-time bottleneck that IS is meant to avoid.
Because $J$ is absent from the proposal, the exact log-weight
$\log w(x) = \log P(x) - \log Q_\mu(x)$ evaluates to
\begin{equation}
  \log w(x)
  = -\mu \sum_i x_i
    + \sum_{(i,j)\in E} J_{ij}\,x_i x_j
    + C,
  \label{eq:is_logweight}
\end{equation}
where $C$ is a sample-independent constant that cancels in the
self-normalised estimator.
The coupling term is essential. If it is omitted, co-failure states are 
consistently underweighted. These are exactly the states that trigger 
the threshold rules and dominate the tail, so $\hat{p}_\mathrm{tail}$ 
remains biased even as $N$ increases.
Samples drawn from $Q_\mu$ are reweighted by $w(x) = \exp(\log w(x))$.
Because $C$ in Eq.~\eqref{eq:is_logweight} is dropped rather than tracked
exactly, $w(x)$ is known only up to the same missing sample-independent
constant, so it cannot be used as a standard (normalised) importance
weight; instead we use the \emph{self-normalised} importance-sampling
estimator,
\begin{equation}
  \hat{p}_{\mathrm{tail}}^{\mathrm{FIS}}
  = \frac{\displaystyle\sum_{k=1}^{N}
    w(x^{(k)})\,\mathbf{1}\bigl[C(x^{(k)}) > T\bigr]}
    {\displaystyle\sum_{k=1}^{N} w(x^{(k)})},
  \label{eq:is_est}
\end{equation}
in which the unknown constant cancels between numerator and denominator
regardless of its value.  This estimator is \emph{consistent}
($\hat{p}_{\mathrm{tail}}^{\mathrm{FIS}} \to p_{\mathrm{tail}}$ as
$N\to\infty$) but not exactly unbiased at finite $N$. Self-normalised
importance sampling carries an $O(1/N)$ bias from the ratio of two random
sums, which shrinks with $N$ but is not identically zero, unlike the
standard (non-self-normalised) IS estimator that would result from
knowing $C$ exactly.  We report this finite-sample offset empirically in
Sec.~\ref{sec:results} rather than assume it away.
The shift $\mu^*$ is calibrated by matching the expected number of
failures under $Q_\mu$ to a fixed target count near the threshold,
not by minimising variance directly.
For \ac{scm20}, $\mu^*\approx1.90$ gives an expected $3.5$ failures at
$T_\mathrm{ref}$, and we reuse this shift at $T_\mathrm{struct}$. The latter
threshold can be crossed when only two specific failures occur
(Sec.~\ref{sec:oracle_struct}), so increasing a global proposal shift can
move probability mass away from the narrow, rule-specific configurations
that matter.

\subsubsection{Iterative Cross-Entropy}
\label{sec:ice}

The iterative Cross-Entropy (\ac{ice}) method~\cite{Rubinstein:1999,Kroese:2013}
generalises fixed-shift IS by \emph{adaptively} learning the proposal
distribution from the data, rather than requiring $\mu^*$ to be set in advance.
The method alternates between a pilot phase, in which the proposal is
refined, and a final estimation phase.

In the pilot phase, a sequence of $n_r$ rounds is run.
At round $r$, a batch of $N_{\mathrm{pilot}}$ samples is drawn from
the current proposal $Q^{(r)}_\mu$, and the proposal parameter is
updated by minimising the Kullback--Leibler divergence from the
conditional target distribution $P(\cdot \mid C > T)$,
\begin{equation}
  \mu^{(r+1)} = \operatorname*{arg\,min}_{\mu}\;
    \KL{P(\cdot \mid C>T)}{Q_\mu}.
  \label{eq:ice_update}
\end{equation}
This update has a closed form for the exponential-tilting family: it
amounts to choosing $\mu$ so that the weighted mean number of failures
under $Q_\mu$ matches the weighted mean under the current pilot
samples restricted to $\{C > T\}$.
Each round steers the proposal closer to the optimal IS distribution,
progressively concentrating mass on the configurations that contribute
most to the tail probability.

After $n_r$ pilot rounds, a final large batch of $N_{\mathrm{final}}$
samples is drawn from the converged proposal $Q^{(n_r)}_\mu$ and the
estimator~\eqref{eq:is_est} is applied.
For the \ac{scm20} we use $n_r = 6$ pilot rounds, each of size
$N_{\mathrm{pilot}} = N/(3n_r)$, followed by a final estimation batch
of $N_{\mathrm{final}} = 2N/3$ samples.
All six pilot rounds together consume $N/3$ samples and the final
estimation the remaining $2N/3$, so the total budget is exactly $N$,
on equal footing with every other classical method.
Because the final proposal is adapted directly to the target tail
region, \ac{ice} is expected to achieve lower variance than fixed-shift
\ac{fixedis} at the same total sample budget, provided the pilot rounds
are sufficient for the proposal to converge~\cite{Papaioannou:2019}.
The actual variance reduction relative to the other classical methods
is reported in Sec.~\ref{sec:results}.

\subsection{Quantum Amplitude Estimation Pipeline}
\label{sec:qae}

This section develops the quantum estimation pipeline used throughout
the paper.  \ac{qae} encodes the tail probability $p_\mathrm{tail}$
defined in Sec.~\ref{sec:cost} as the squared amplitude of an ancilla
qubit. The amplitude-encoding operator $\mathcal{A}$ decomposes into two
independent building blocks: a state-preparation unitary $\mathcal{R}$
and a threshold oracle $\mathcal{F}_T$. Their specific constructions,
developed in Secs.~\ref{sec:state_prep} and~\ref{sec:oracle_design},
constitute the paper's two central methodological contributions. This
section defines only the general structure that both components must
satisfy. We then introduce \ac{mlae}, the practical, \ac{qpe}-free protocol used to
extract the amplitude from repeated circuit executions, and the
shallow-circuit rationale for preferring it over the canonical
algorithm.

\subsubsection{The \texorpdfstring{$\mathcal{A}$}{A} Operator and Amplitude Encoding}
\label{sec:qae_operator}

Quantum Amplitude Estimation targets an unknown probability $a \in [0,1]$
encoded as the squared amplitude of a designated ancilla qubit.
The starting point is a unitary $\mathcal{A}$ acting on an $(n+1)$-qubit
register such that
\begin{equation}
  \mathcal{A}\ket{0}^{\otimes(n+1)}
  = \sqrt{1-a}\,\ket{\Psi_0}\ket{0} + \sqrt{a}\,\ket{\Psi_1}\ket{1},
  \label{eq:A_decomp}
\end{equation}
where the last qubit is the ancilla and $\ket{\Psi_0}$, $\ket{\Psi_1}$ are
normalised states of the $n$-qubit data register.
Measuring the ancilla yields $\ket{1}$ with probability exactly $a$.

In our supply-chain risk setting, $\mathcal{A}$ is constructed from two
building blocks.
The first is the \emph{state-preparation unitary} $\mathcal{R}$, which
prepares an approximation $Q^*$ to the Boltzmann distribution $P(x)$ on
the data register:
\begin{equation}
  \mathcal{R}\ket{0}^{\otimes n}
  = \sum_{x \in \{0,1\}^n} \sqrt{Q^*(x)}\,\ket{x}.
  \label{eq:R_def}
\end{equation}
The specific construction of $\mathcal{R}$ used in this work is presented
in Sec.~\ref{sec:state_prep}.
The second building block is the \emph{threshold oracle} $\mathcal{F}_T$,
which marks every configuration whose disruption cost exceeds the threshold
$T$ by flipping the ancilla qubit:
\begin{equation}
  \mathcal{F}_T\ket{x}\ket{0}
  = \ket{x}\ket{\mathbf{1}[C(x) > T]}.
  \label{eq:oracle_def}
\end{equation}
Two implementations of $\mathcal{F}_T$ are presented in
Sec.~\ref{sec:oracle_design}.
The combined operator is
\begin{equation}
  \mathcal{A} = \mathcal{F}_T \circ \bigl(\mathcal{R} \otimes I\bigr).
\end{equation}
Applying $\mathcal{A}$ to $\ket{0}^{\otimes(n+1)}$ and tracing out the
data register, the probability of observing $\ket{1}$ on the ancilla is
\begin{equation}
  a = \sum_{x:\,C(x) > T} Q^*(x) = \mathbb{E}_{Q^*}[\mathbf{1}[C(x)>T]],
\end{equation}
the tail probability under $Q^*$.
When $Q^*$ is a good approximation to $P$, $a \approx P(C > T)$.

The structure of $\mathcal{A}$ implies that the full $(n+1)$-qubit Hilbert
space contains a 2-dimensional invariant subspace spanned by
$\ket{\Psi_0}\ket{0}$ and $\ket{\Psi_1}\ket{1}$, within which amplitude
amplification operates regardless of the $2^n$ complexity of the full space.

\subsubsection{Maximum Likelihood Amplitude Estimation}
\label{sec:mlae}

Writing $a = \sin^2\!\theta_a$, the state~\eqref{eq:A_decomp} becomes
$\cos\theta_a\,\ket{\Psi_0}\ket{0} + \sin\theta_a\,\ket{\Psi_1}\ket{1}$.
The Grover operator is
\begin{equation}
  \mathcal{Q} = -\mathcal{A}\,S_0\,\mathcal{A}^\dagger\,S_f,
  \label{eq:grover_operator}
\end{equation}
where $S_0 = I - 2\ket{0}^{\otimes(n+1)}\bra{0}^{\otimes(n+1)}$ is the
reflection about the all-zero computational basis state and
$S_f = I - 2(I_\mathrm{data}\otimes\ket{1}\bra{1})$
flips the phase of the tail component.
The leading $-1$ is a global phase that does not affect measurement
probabilities and is omitted in the implementation.
The product of these two reflections acts as a rotation by $2\theta_a$
within the invariant two-dimensional subspace spanned by
$\ket{\Psi_0}\ket{0}$ and $\ket{\Psi_1}\ket{1}$.
After $k$ applications, $\mathcal{Q}^k\mathcal{A}\ket{0}^{\otimes(n+1)}$
places the ancilla amplitude at angle $(2k+1)\theta_a$, so measuring
$\ket{1}$ on the ancilla has probability
\begin{equation}
  p_k = \sin^2\!\bigl((2k+1)\theta_a\bigr).
  \label{eq:pk}
\end{equation}
The canonical \ac{qae} algorithm uses \ac{qpe} to
extract $\theta_a$ from the eigenvalues $e^{\pm 2i\theta_a}$ of
$\mathcal{Q}$~\cite{Brassard:2002}.
\ac{qpe} achieves estimation error $\epsilon$ with $O(1/\epsilon)$ oracle calls,
but requires $O(\log(1/\epsilon))$ additional ancilla qubits and
controlled-$\mathcal{Q}^{2^j}$ gates whose depth grows exponentially with
circuit precision, making it impractical on near-term hardware.

\ac{mlae}~\cite{Suzuki:2020} achieves
the same asymptotic scaling with shallow circuits and no additional
ancilla, provided the schedule's maximum depth grows with the shot
budget (Sec.~\ref{sec:results_schedule_scaling}).
The circuit $\mathcal{Q}^k\mathcal{A}$ is executed for each depth $k$ in a
fixed schedule $\mathcal{K} = \{k_0, k_1, \ldots\}$, yielding $h_k$
observations of $\ket{1}$ out of $N_k$ shots.
Each outcome is a Bernoulli draw with success probability $p_k$
given by~\eqref{eq:pk}, so the log-likelihood is
\begin{equation}
  \ell(\theta_a) = \sum_{k \in \mathcal{K}}
  \Bigl[h_k \log \sin^2\!\bigl((2k+1)\theta_a\bigr) 
        + (N_k - h_k) \log \cos^2\!\bigl((2k+1)\theta_a\bigr)\Bigr].
\end{equation}
The \ac{mlae} estimator $\hat{\theta}_a = \arg\max_{\theta_a}\ell(\theta_a)$
is found by one-dimensional numerical optimisation, and the amplitude
estimate is $\hat{a} = \sin^2\!\hat{\theta}_a$.
The total shot budget consumed is
\begin{equation}
  N = \sum_{k \in \mathcal{K}} N_k = N_\text{shots/k} \times |\mathcal{K}|.
\end{equation}
In this work we use the schedule $\mathcal{K} = \{0, 1, 2, 3\}$ with equal
shots per depth, following~\cite{Suzuki:2020}.
For a schedule whose maximum depth grows with the total query budget,
\ac{mlae}'s mean squared error scales as $O(N_\mathrm{query}^{-2})$, compared
to $O(N_\mathrm{query}^{-1})$ for classical Monte Carlo, corresponding to a
quadratic speedup
in the number of oracle queries required to achieve a target precision
$\epsilon$: $O(1/\epsilon)$ versus $O(1/\epsilon^2)$.
This ideal rate is not what the \emph{fixed}, bounded-depth schedule used
throughout this paper achieves. Sec.~\ref{sec:results_schedule_scaling}
shows analytically and empirically that any fixed schedule, however deep,
yields only the classical $O(N_\mathrm{query}^{-1})$ rate, with depth
improving the constant factor rather than the asymptotic exponent.

\subsection{State Preparation via Belief Propagation and Chow-Liu Trees}
\label{sec:state_prep}

\subsubsection{Chow-Liu Tree Approximation}

The state-preparation unitary $\mathcal{R}$ was defined in
Eq.~\eqref{eq:R_def}; its goal is to approximate the Ising
Boltzmann distribution in Eq.~\eqref{eq:boltzmann}.
In general, this distribution contains correlations induced by the
full interaction graph $E$.
Preparing it exactly can therefore be costly.
A useful approximation is to replace $P$ by a distribution $Q$ whose
dependency structure is a tree.
Such a distribution can be factorised and implemented more efficiently
while retaining the strongest pairwise dependencies present in~$P$.

The Chow--Liu theorem~\cite{ChowLiu:1968} gives the optimal
tree-structured approximation to $P$ in the sense of minimising the
Kullback--Leibler (KL) divergence $D_{\mathrm{KL}}(P\,\|\,Q)$.
The key quantity is the pairwise mutual information (MI) between
variables $X_i$ and $X_j$,
\begin{equation}
  I(X_i;X_j)
  = \sum_{x_i,x_j} P(x_i,x_j)
    \log\frac{P(x_i,x_j)}{P(x_i)P(x_j)}.
  \label{eq:mi_def}
\end{equation}
This measures the statistical dependence between two variables; larger
values indicate stronger pairwise correlations.

For a fixed spanning tree $\mathcal{T}$ (a cycle-free subgraph connecting
all vertices), let $Q_\mathcal{T}$ denote the
tree-structured distribution whose one- and two-variable marginals match
those of $P$ on the vertices and edges of $\mathcal{T}$.
Its factorisation is
\begin{equation}
  Q_\mathcal{T}(x)
  = P(x_r)\prod_{i\neq r} P\!\bigl(x_i \mid x_{\pi(i)}\bigr),
  \label{eq:tree_factorisation}
\end{equation}
where $r$ is an arbitrary root node and $\pi(i)$ denotes the parent
of node $i$ in the rooted tree.
Denoting by
\begin{equation}
  H(X_i) = -\sum_{x_i} P(x_i)\log P(x_i)
  \label{eq:marginal_entropy}
\end{equation}
the marginal Shannon entropy of node $i$, the KL divergence between
$P$ and this tree approximation satisfies
\begin{equation}
  D_{\mathrm{KL}}(P\,\|\,Q_\mathcal{T})
  = \sum_i H(X_i)
    - \sum_{(i,j)\in\mathcal{T}} I(X_i;X_j)
    - H(P),
  \label{eq:kl_tree}
\end{equation}
where $H(P)$ is the joint entropy of the full distribution.
Since $\sum_i H(X_i)$ and $H(P)$ are both independent of the choice
of $\mathcal{T}$, minimising the KL divergence is equivalent to
maximising the total mutual information over the tree edges:
\begin{equation}
  \mathcal{T}^*
  = \arg\max_{\mathcal{T}}
    \sum_{(i,j)\in\mathcal{T}} I(X_i;X_j).
  \label{eq:chow_liu_objective}
\end{equation}

Thus, the optimal Chow--Liu tree $\mathcal{T}^*$ is the maximum-weight
spanning tree of the complete graph whose edge weights are the pairwise
mutual information values --- equivalently, the minimum spanning tree on
$-I(X_i;X_j)$, found in $O(n^2\log n)$ by Kruskal's greedy
algorithm~\cite{kruskal:1956}.
The resulting approximation is
\begin{equation}
  Q^*(x)
  = P(x_r)\prod_{i\neq r} P\!\bigl(x_i \mid x_{\pi(i)}\bigr),
  \label{eq:chow_liu_factorisation}
\end{equation}
where the parent map $\pi(i)$ is now determined by $\mathcal{T}^*$.
This distribution preserves the most informative pairwise dependencies
of the original Ising model while reducing the dependency structure
to a tree.

Because the graph has four disconnected components, its 19-edge
Chow--Liu tree cannot be formed simply by dropping edges. The tree
retains 16 of the original 20 edges: 14 from a spanning tree of the
core and the single edge from each two-node component. It drops the
remaining 4 core edges and adds 3 edges that are absent from the
original Ising graph to connect the four components. Despite these
changes, the Chow--Liu approximation remains highly accurate. The
4 dropped edges have mutual information below
$2\times10^{-4}$ nats, while the 3 added edges connect independent
components and therefore carry exactly zero true mutual information.
The circuit distribution retains approximately $83\%$ of the exact
pairwise covariance at the chosen threshold.

\subsubsection{Belief Propagation for Ising Marginals}

Constructing $Q^*$ requires only the one-body marginals
$p_i = P(x_i = 1)$ and two-body marginals
$p_{ij} = P(x_i=1, x_j=1)$;
the MI matrix and the conditional rotation angles
in~\eqref{eq:chow_liu_factorisation} follow from these
$O(n^2)$ numbers alone.

We compute these marginals via \ac{bp}~\cite{Yedidia:2003, Murphy:1999, Aji:2000}, a
message-passing algorithm on the Ising factor graph.
Equation~\eqref{eq:boltzmann} factorises as
\begin{equation}
  P(x) \propto
  \prod_i \phi_i(x_i)
  \prod_{(i,j)\in E}\psi_{ij}(x_i,x_j),
  \label{eq:ising_factorisation}
\end{equation}
with unary factors $\phi_i(x_i) = e^{h_i x_i}$
and pairwise factors $\psi_{ij}(x_i,x_j) = e^{J_{ij}x_ix_j}$,
obtained by splitting the exponential of a sum into a product
of exponentials.
\ac{bp} assigns a directed message $\mu_{i\to j}(x_j)$ to every 
edge and updates it as
\begin{equation}
  \mu_{i\to j}(x_j)
  = \sum_{x_i\in\{0,1\}}
    \phi_i(x_i)\,\psi_{ij}(x_i,x_j)
    \!\!\prod_{k\in\partial i\setminus j}\!\!
    \mu_{k\to i}(x_i),
  \label{eq:bp_update}
\end{equation}
where $\partial i\setminus j$ denotes the neighbours of $i$
excluding $j$.
Messages are initialised uniformly, $\mu_{i\to j}^{(0)}(x_j)=\tfrac{1}{2}$ for 
all $x_j$, and then updated synchronously (all directed messages in parallel) 
in log-domain to avoid numerical underflow~\cite{KollerFriedman:2009}.
After convergence the one-body marginal is
\begin{equation}
  p_i
  = \sigma\!\Bigl(h_i
    + \sum_{k\in\partial i}
      \log\frac{\mu_{k\to i}(1)}{\mu_{k\to i}(0)}\Bigr),
  \label{eq:bp_marginal}
\end{equation}
where $\sigma(z)=(1+e^{-z})^{-1}$ is the logistic function.
Two-body marginals on Ising edges are extracted from the
converged pairwise beliefs using cavity log-weights (a cavity
log-weight aggregates a node's incoming messages from every
neighbour except one, as if that neighbour's edge had been removed).
For non-edge pairs the independence approximation
$p_{ij}\approx p_i p_j$ is used; for pairs in different connected
components this is exact, since the Ising factorisation
(Eq.~\eqref{eq:ising_factorisation}) has no interaction term linking
them, so their true mutual information is exactly zero.  
For non-adjacent pairs \emph{within} the same component, this is only an
approximation and is not guaranteed a priori to be negligible. In
general, Ising variables can carry non-zero mutual information through
paths in the graph.
The spanning tree $\mathcal{T}^*$ actually
constructed is therefore the Chow--Liu optimum with respect to this
approximate mutual-information matrix, not a proven optimum with
respect to the exact $P$.  We do not assume the approximation away;
instead, Sec.~\ref{sec:results} validates the resulting circuit
distribution directly against the exact Boltzmann distribution $P$ by
full enumeration, so its practical impact is measured empirically for
\ac{scm20} rather than asserted analytically.
Full derivation of~\eqref{eq:bp_update}--\eqref{eq:bp_marginal}
and the pairwise belief computation is given in
Appendix~\ref{app:bp}.

\ac{bp} gives exact marginals on trees; for graphs with $L$
independent cycles it is approximate but typically
accurate for sparse, weakly-coupled
graphs~\cite{Murphy:1999}.
The \ac{scm20} Ising graph has cyclomatic number $L = 4$, but these
cycles are short and confined to small, low-treewidth blocks of the
fifteen-node core component rather than forming one large loop, so
\ac{bp} remains a good approximation in this sparse, weakly-coupled
regime.
\ac{bpacl} requires zero samples, so the entire shot budget is
available for \ac{mlae} (Sec.~\ref{sec:budget}).

\subsubsection{Quantum Circuit Construction}

The Chow--Liu factorisation in Eq.~\eqref{eq:chow_liu_factorisation}
is encoded into a quantum circuit $\mathcal{R}$ using the
conditional-rotation construction for quantum Bayesian networks
introduced in Refs.~\cite{Low:2014,Borujeni:2021}. In our setting,
the circuit topology follows the directed Chow--Liu tree, while the
rotation angles are computed from the \ac{bp} marginals rather than from
exact conditional probability tables.

Let $r$ denote the root node. The root qubit is initialised by applying
\begin{equation}
  R_y(\theta_r),
  \qquad
  \theta_r = 2\arcsin\!\sqrt{p_r},
  \label{eq:root_ry}
\end{equation}
so that
\begin{equation}
  \bigl|\bra{1}R_y(\theta_r)\ket{0}\bigr|^2 = p_r ~,~ \bigl|\langle 0|R_y(\theta_r)|0\rangle\bigr|^2 = 1 - p_r.
\end{equation}
Thus, the marginal probability of the root node is encoded directly
into the amplitude of qubit $r$.

For each non-root node $i$ with parent $\pi(i)$, we first compute the
conditional probabilities from the \ac{bp} marginals. 
Writing $p_{i,\pi(i)}$ for the two-body marginal $p_{ij}$ at the parent 
edge $j=\pi(i)$, the conditional probabilities are
\begin{align}
  P\!\bigl(x_i=1\mid x_{\pi(i)}=1\bigr)
  &= \frac{p_{i,\pi(i)}}{p_{\pi(i)}},
  \label{eq:cond_prob_1}\\
  P\!\bigl(x_i=1\mid x_{\pi(i)}=0\bigr)
  &= \frac{p_i - p_{i,\pi(i)}}{1 - p_{\pi(i)}}.
  \label{eq:cond_prob_0}
\end{align}
The corresponding conditional rotation angles are
\begin{equation}
  \theta_{i|s}
  =
  2\arcsin\!\sqrt{P(x_i=1\mid x_{\pi(i)}=s)},
  \qquad s\in\{0,1\}.
  \label{eq:conditional_theta}
\end{equation}
These are decomposed into the uniformly controlled rotation parameters
\begin{equation}
  \alpha_i = \frac{\theta_{i|0}+\theta_{i|1}}{2},
  \qquad
  \beta_i = \frac{\theta_{i|0}-\theta_{i|1}}{2}.
  \label{eq:alpha_beta}
\end{equation}

The conditional distribution of node $i$ is then implemented using the
two-CNOT uniformly controlled block
\begin{equation}
  \mathrm{CX}\bigl(\pi(i)\!\to\!i\bigr)
  -\, R_y(\beta_i)
  -\, \mathrm{CX}\bigl(\pi(i)\!\to\!i\bigr)
  -\, R_y(\alpha_i).
  \label{eq:cond_ry}
\end{equation}
When the parent qubit is in basis state $\ket{s}$, the net rotation
applied to the child qubit is $R_y(\theta_{i|s})$. Hence, the block
encodes the conditional probability
$P(x_i=1\mid x_{\pi(i)}=s)$ for both parent states $s\in\{0,1\}$.

The full circuit is constructed by applying the root rotation
\eqref{eq:root_ry}, followed by the conditional block
\eqref{eq:cond_ry} for every non-root node in a parent-before-children
ordering of the Chow--Liu tree rooted at $r$. By linearity,
the same conditional rotations act coherently when parent qubits are
in superposition, so the resulting state prepares the joint
distribution specified by Eq.~\eqref{eq:chow_liu_factorisation}.
A formal proof is given in Appendix~\ref{app:chow_liu}.

For the model considered here, with $n=20$, the circuit contains
$2n-1=39$ single-qubit $R_y$ gates (one for the root, two for each of
the $n-1$ non-root nodes) and $2(n-1)=38$ CNOT gates, has total depth
$43$, and requires no ancilla qubits. The circuit topology is
the same whether the marginals are obtained from \ac{bp} or \ac{mcmc}; only the
rotation angles $\alpha_i$ and $\beta_i$ change.

\subsection{Oracle Design}
\label{sec:oracle_design}

The threshold oracle $\mathcal{F}_T$ is the second building block
of the $\mathcal{A}$ operator (Sec.~\ref{sec:qae_operator}),
implementing Eq.~\eqref{eq:oracle_def} on the $n$-qubit data
register and a single ancilla qubit, where $C(x)$ is the total
disruption cost defined in Sec.~\ref{sec:model}.

We have used a model oracle $\mathcal{F}_T^\mathrm{model}$
as an exact reference. This oracle
evaluates $C(x)$
classically for every $x \in \{0,1\}^n$ and constructs an
$(n+1)$-qubit diagonal unitary $D_{n+1}$, acting jointly on the data
register and the ancilla qubit $q$, such that
\begin{equation}
  D_{n+1}\ket{x}\ket{q}
  =
  \begin{cases}
  \ket{x}\ket{q} & q = 0, \\[2pt]
  (-1)^{\mathbf{1}[C(x) > T_\mathrm{ref}]}\ket{x}\ket{q} & q = 1,
  \end{cases}
  \label{eq:diagonal_gate}
\end{equation}
so the phase flip only ever applies on the ancilla-one branch, and
only for scenarios above threshold. The ancilla starts in $\ket{0}$
and is first mapped to $\ket{+} = (\ket{0}+\ket{1})/\sqrt{2}$ by a
Hadamard gate. Applying $D_{n+1}$ then leaves the $\ket{0}$ branch of
the ancilla unchanged and multiplies its $\ket{1}$ branch by $-1$
exactly for the tail scenarios. A second Hadamard on the ancilla
turns this conditional phase into a conditional bit flip, giving the
full construction
\begin{equation}
  \mathcal{F}_T^\mathrm{model}
  =
  (I_n \otimes H)\, D_{n+1} \,(I_n \otimes H),
  \label{eq:model_oracle_construction}
\end{equation}
which implements Eq.~\eqref{eq:oracle_def} exactly, by the phase
kickback principle~\cite{Woerner:2019}.
The construction is exact for any threshold $T_\mathrm{ref}$ and
requires $n + 1$ qubits.
However, classical enumeration costs $O(n \cdot 2^n)$ operations
and a general diagonal unitary on $n+1$ qubits decomposes into
$O(2^n)$ elementary two-qubit gates~\cite{Shende:2006}, making
the circuit intractable on a \ac{qpu}.
The model oracle therefore serves exclusively as an exact
simulation reference for validating the structural oracle and
quantifying the approximation error introduced by both the
Chow-Liu state preparation and the structural threshold.

\subsubsection{Structural Oracle}
\label{sec:oracle_struct}

The structural oracle $\mathcal{F}_T^\mathrm{struct}$ avoids the
$2^n$ enumeration entirely by exploiting the additive Boolean
structure of $C(x)$.
It is composed exclusively of $X$, CNOT, Toffoli, and 3-controlled-$X$ gates,
and requires $n + 8$ qubits. Its gate count does not grow exponentially
with $n$, unlike the model oracle, but the resulting circuit is still
costly. Sec.~\ref{sec:budget_gates} reports the exact gate counts.
Its design is the primary contribution of this section and is
detailed below.

The total disruption cost (Sec.~\ref{sec:model}) decomposes as
\begin{equation}
  C(x)
  = \sum_i c_i x_i
  + \sum_{(i,j)\in\mathcal{P}} \pi_{ij} x_i x_j
  + \sum_{r=1}^{2} \Delta_r\,\mathbf{1}[\mathrm{rule}_r(x)],
  \label{eq:cost_decomp}
\end{equation}
where the first two terms are linear and pairwise costs, and
each $\Delta_r$ is a large penalty activated when the Boolean
rule $\mathrm{rule}_r(x)$ fires.
The model contains two such rules:
\begin{align}
  \mathrm{rule}_1(x)
  &= \mathbf{1}\!\Bigl[\sum_{i \in G_1} x_i \geq k\Bigr],
  \label{eq:rule_count}\\
  \mathrm{rule}_2(x)
  &= \mathbf{1}\!\bigl[
     \mathrm{any}(x_{G_{2A}}) \wedge \mathrm{any}(x_{G_{2B}})
     \bigr],
  \label{eq:rule_and_any}
\end{align}
where $G_1$ is a group of tier-1 supplier variables with
failure-count threshold $k$,
$G_{2A}$ and $G_{2B}$ are plant and port variable groups,
and $\Delta_1$, $\Delta_2$ are cost penalties large enough
that any rule-firing state exceeds any reasonable tail
threshold.

The structural oracle does not compute the full cost $C(x)$.
Instead, it chooses a threshold for which every state that activates
one of the penalty rules is guaranteed to lie in the tail. This makes
it sufficient for the oracle to test only whether a rule fires.

For each rule $r$, let $C_\mathrm{min}^{(r)}$ denote a lower bound on
the smallest total cost among all states that activate that rule,
computed from the relevant linear costs and the rule penalty alone.
All costs and penalties in the model are nonnegative, so omitting the
pairwise term can only understate the true cost. $C_\mathrm{min}^{(r)}$
is therefore a valid lower bound, not necessarily the exact minimum.
For a count-threshold rule on a group $G_r$, where the rule fires
whenever at least $k_r$ variables in $G_r$ are equal to one, this
bound is
\begin{equation}
  C_\mathrm{min}^{(r)}
  =
  \sum_{i=1}^{k_r} c_{(i)}^{(r)} + \Delta_r,
  \label{eq:cmin_count}
\end{equation}
where
$c_{(1)}^{(r)} \leq c_{(2)}^{(r)} \leq \cdots$
are the sorted linear costs of the variables in $G_r$.
This is a lower bound on the cost of the cheapest activating state,
since it omits any pairwise penalty among the chosen failures.

For an and-any rule involving two groups $G_A$ and $G_B$, the rule
fires when at least one variable in each group is equal to one. The
corresponding lower bound is
\begin{equation}
  C_\mathrm{min}^{(r)}
  =
  \min_{i \in G_A} c_i
  +
  \min_{j \in G_B} c_j
  +
  \Delta_r.
  \label{eq:cmin_and}
\end{equation}
As before, this omits any pairwise penalty between the two chosen
variables.

We define the structural threshold as one unit below the smallest of
these lower bounds over all rules:
\begin{equation}
  T_\mathrm{struct}
  =
  \min_r C_\mathrm{min}^{(r)} - 1.
  \label{eq:t_struct}
\end{equation}
With this choice, any state that activates any rule necessarily exceeds
the structural threshold. Indeed, for every rule $r$,
\begin{equation}
  \mathrm{rule}_r(x) = 1
  \;\Longrightarrow\;
  C(x) \geq C_\mathrm{min}^{(r)} > T_\mathrm{struct}.
  \label{eq:rule_implication}
\end{equation}

Therefore, at the structural threshold, the oracle can mark states by
evaluating only the Boolean predicate
\begin{equation}
  \mathrm{rule}_1(x) \vee \mathrm{rule}_2(x),
\end{equation}
rather than evaluating the full disruption cost $C(x)$.
This replacement is exact for all rule-firing states. Every such state
is marked and belongs to the event $C(x) > T_\mathrm{struct}$.

The approximation arises only from states that do not activate any
rule but whose linear and pairwise costs alone exceed
$T_\mathrm{struct}$. These residual tail states satisfy
$C(x) > T_\mathrm{struct}$ but are not marked by
$\mathcal{F}_T^\mathrm{struct}$.

Because $g(x) = \mathrm{rule}_1(x) \vee \mathrm{rule}_2(x)$ marks a
strict subset of the tail event, the quantity actually estimated on the
structural-oracle path is not $p_\mathrm{tail} = P(C(x)>T_\mathrm{struct})$
itself, but the probability of a \emph{surrogate} event,
\begin{equation}
  p_\mathrm{struct} = P\bigl(g(x)=1\bigr).
  \label{eq:surrogate_def}
\end{equation}
The implication~\eqref{eq:rule_implication} gives
$\{x : g(x)=1\} \subseteq \{x : C(x) > T_\mathrm{struct}\}$, so
$p_\mathrm{struct}$ is always a rigorous \emph{lower bound} on the true
tail probability. Defining the missed probability mass
\begin{equation}
  \epsilon_\mathrm{miss}
  = p_\mathrm{tail} - p_\mathrm{struct}
  = P\bigl(C(x) > T_\mathrm{struct} \wedge \neg g(x)\bigr)
  \;\geq\; 0
  \label{eq:eps_miss}
\end{equation}
gives the elementary two-sided bound
\begin{equation}
  p_\mathrm{struct}
  \;\leq\;
  p_\mathrm{tail}
  \;=\;
  p_\mathrm{struct} + \epsilon_\mathrm{miss}.
  \label{eq:surrogate_bound}
\end{equation}
Every quantity reported for the structural-oracle path --- both the
\ac{mlae} amplitude and the classical baselines using the same oracle
--- therefore estimates $p_\mathrm{struct}$, a certified lower bound on
the true tail probability, and not $p_\mathrm{tail}$ itself. The
resulting systematic under-marking error $\epsilon_\mathrm{miss}$ is
bounded empirically against the model-oracle reference and discussed
in Sec.~\ref{sec:results}.

The gate-level construction of $\mathcal{F}_T^\mathrm{struct}$
--- three reversible sub-circuits (OR gadget, ripple-carry counter,
and comparator) composed via uncomputation --- is detailed in
Appendix~\ref{app:oracle_circuit}.
The net action on the $n$-qubit data register and eight ancilla
qubits is
\begin{equation}
  \mathcal{F}_T^\mathrm{struct}
  : \ket{x}\ket{0}^{\otimes 8}
  \;\mapsto\;
  \ket{x}
  \ket{\mathbf{1}[\mathrm{rule}_1(x) \vee \mathrm{rule}_2(x)]}
  \ket{0}^{\otimes 7}.
  \label{eq:struct_oracle_action}
\end{equation}
The data register is unchanged, all scratch qubits return to
$\ket{0}$, and no classical enumeration of $C(x)$ is required. The
circuit is built entirely from standard gates, though its resource
cost places it in the fault-tolerant regime rather than the near-term
one. This construction is a first, unoptimised implementation. Future
work could reduce its gate count or explore alternative oracle designs
for the same class of threshold rules.

\subsubsection{Error Decomposition}
\label{sec:error_decomposition}

The preceding sections introduce two independent, deterministic
approximations: the Chow--Liu state $Q^*$ in place of the exact
Boltzmann distribution $P$ (Sec.~\ref{sec:state_prep}), and the
surrogate oracle event $g(x)$ in place of the exact tail indicator
(Eq.~\eqref{eq:surrogate_def}, this section).  Together with the
statistical estimation error of \ac{mlae} itself, these are the only
three sources of error in a reported tail-probability estimate, and
they combine additively.  Writing $\hat{a}$ for the \ac{mlae} point
estimate of the amplitude $a$ (Sec.~\ref{sec:mlae}) and $P(C>T)$ for
the exact ground truth, the exact identity
\begin{equation}
  \hat{a} - P(C>T)
  =
  \underbrace{(\hat{a} - a)}_{\text{statistical}}
  \,+\,
  \underbrace{(a - Q^*(C>T))}_{\text{oracle error}}
  \,+\,
  \underbrace{(Q^*(C>T) - P(C>T))}_{\text{state-preparation bias}}
  \label{eq:error_decomposition}
\end{equation}
separates the total error into three terms, each characterised
separately elsewhere in the paper.  The statistical term is the only
one that shrinks with the shot budget. Its root-mean-square error
scales as $O(N_\mathrm{query}^{-1/2})$ under the fixed schedule used
throughout this paper (Sec.~\ref{sec:results_query}), approaching the
ideal $O(N_\mathrm{query}^{-1})$ only if the schedule's maximum depth is
allowed to grow with the budget (Sec.~\ref{sec:results_schedule_scaling}).
The oracle-error term is deterministic, independent of the shot budget,
zero by construction for the model oracle, and, in the $Q^*\to P$
limit, coincides with $-\epsilon_\mathrm{miss}$
(Eq.~\eqref{eq:eps_miss}).  The state-preparation-bias term is likewise
deterministic and independent of which oracle is used.  No increase in
\ac{mlae} shots can reduce either of the two deterministic terms; only a
better Chow--Liu approximation or a tighter oracle construction can.
Sec.~\ref{sec:results} reports the numerical value of every term in
Eq.~\eqref{eq:error_decomposition} for both thresholds studied.

\subsection{Resource Accounting: Shots, Queries, and Gate-Level Cost}
\label{sec:budget}

A meaningful comparison between quantum and classical estimators requires
a common unit of computational cost.  Rather than defend a single metric,
we report three complementary resource views, following the principle that
different abstractions expose different aspects of cost: the number of
circuit executions or samples (statistical information per completed
run), the number of amplitude-encoding queries (the standard
complexity-theoretic unit for amplitude estimation), and gate-level
circuit resources (implementation feasibility on real hardware).  We
define each below and use all three in Sec.~\ref{sec:results}.

\subsubsection{Shots and Samples}

We define the \emph{sample budget} $N$ as the total number of independent
random draws or circuit executions consumed to produce a \emph{single}
estimate of $P(C > T)$, not counting deterministic preprocessing.

For the classical methods, the budget is straightforward.
\ac{naivemc} draws $N$ i.i.d.\ samples from the product marginals, so
its budget is $N$ by definition. \ac{fixedis} draws $N$ samples from a
biased Bernoulli proposal. The shift $\mu^*$ is computed directly from
the model parameters~\cite{Rubinstein:2016} and consumes no random
samples.

\ac{gibbsmc} runs a Markov chain for
$N_\mathrm{burn}=\max(5\,000,N/2)$ burn-in steps before collecting the
$N$ retained samples used for estimation. These burn-in steps are not
included in the reported budget, so the true total cost is
$N_\mathrm{burn}+N$. The resulting overhead decreases from roughly
$26\times$ at the smallest sweep point ($N=200$) to about $1.5\times$
at the largest ($N=16\,000$).

The value of $N_\mathrm{burn}$ is a fixed, conservative default rather
than one tuned to this model. A pilot run of $10^5$ post-burn-in sweeps
gives an integrated autocorrelation time of $\tau\approx0.5$ for the
\ac{scm20} chain, so $5\,000$ steps substantially exceeds what this
fast-mixing instance requires to equilibrate. We retain the untuned default
for transparency. Autocorrelation affects estimator variance through the
effective sample size
$N_\mathrm{eff}=N/(2\tau)$ (Eq.~\eqref{eq:neff}). At the measured value,
$N_\mathrm{eff}\approx N$, so essentially no statistical power is lost
to autocorrelation. The uncounted burn-in is therefore the dominant
budget discrepancy and gives \ac{gibbsmc} a one-sided advantage over
the i.i.d.\ methods at the same nominal $N$.

\ac{ice} runs $n_r=6$ iterative pilot rounds, each consuming
$N_\mathrm{pilot}=N/(3n_r)$ samples for cross-entropy proposal
adaptation. A final estimation round then consumes
$N_\mathrm{final}=2N/3$ samples. The total allocation is
$n_rN_\mathrm{pilot}+N_\mathrm{final}=N/3+2N/3=N$, so \ac{ice} operates
on the same budget as every other classical method.

For the quantum pipeline, the \emph{shot budget} is
\begin{equation}
  N_\mathrm{shots} = N_k \times |\mathcal{K}|,
  \label{eq:n_shots}
\end{equation}
where $\mathcal{K} = \{0,1,2,3\}$ is the Grover-depth schedule and
$N_k = N_\mathrm{shots}/4$ shots are allocated equally to each depth.
The state-preparation circuit $\mathcal{R}$ is
constructed by the deterministic \ac{bp}~+~Chow-Liu pipeline
(Sec.~\ref{sec:state_prep}), which requires zero random samples, so the
entire shot budget is invested in \ac{mlae} circuit executions with no
hidden preprocessing overhead.  Counted this way, one classical sample and
one quantum shot are both single completed executions that each yield one
Bernoulli observation, which is the sense in which $N_\mathrm{shots}$ is
directly comparable to the classical sample budget $N$.

\subsubsection{Query-Normalised Budget}
\label{sec:budget_query}

Counting shots alone is not, however, a fair \emph{query-complexity}
comparison, because a shot at Grover depth $k$ is not equivalent in cost
to a shot at depth $0$.  The \ac{mlae} circuit executed at depth $k$ is
$\mathcal{Q}^k\mathcal{A}$, where $\mathcal{Q}$ is the Grover operator of
Eq.~\eqref{eq:grover_operator} (Sec.~\ref{sec:mlae}).  Each application of
$\mathcal{Q}$ invokes $\mathcal{A}$ and $\mathcal{A}^\dagger$ once each, so
a single shot at depth $k$ consumes $2k+1$ calls to the amplitude-encoding
operator $\mathcal{A}$.  The natural quantum query budget is therefore
\begin{equation}
  N_\mathrm{query} = \sum_{k \in \mathcal{K}} N_k\,(2k+1),
  \label{eq:n_query}
\end{equation}
which for the equal-shot schedule $\mathcal{K} = \{0,1,2,3\}$ used
throughout this paper evaluates to
\begin{equation}
  N_\mathrm{query}
  = N_k\bigl[(1) + (3) + (5) + (7)\bigr]
  = 4\,N_k \times |\mathcal{K}|
  = 4\,N_\mathrm{shots}.
  \label{eq:n_query_factor}
\end{equation}
Every classical method, by contrast, consumes exactly one evaluation of
the stochastic oracle per sample, so $N_\mathrm{query} = N_\mathrm{shots}
= N$ for \ac{naivemc}, \ac{gibbsmc}, \ac{fixedis}, and \ac{ice}.  A budget
of $N_\mathrm{shots}$ quantum shots is therefore not directly comparable,
query-for-query, to $N_\mathrm{shots}$ classical samples. It corresponds
to $4N_\mathrm{shots}$ amplitude-encoding queries under the schedule used
here.  Sec.~\ref{sec:results} reports both the shot-normalised comparison
(Figs.~\ref{fig:comparison_tref}--\ref{fig:comparison_tstruct}) and the
query-normalised comparison (Fig.~\ref{fig:rmse_query}), so that the
reader can evaluate the reported advantage under either cost model.

\subsubsection{Gate-Level Resources}
\label{sec:budget_gates}

Because one \ac{mlae} circuit execution and one classical function
evaluation are vastly different physical operations, we additionally
report circuit-level resources for the structural oracle: the number of
CX, CCX (Toffoli), and multi-controlled-$X$ gates and the circuit depth
of $\mathcal{Q}^k\mathcal{A}$ at each Grover depth $k \in \mathcal{K}$,
obtained by transpiling the circuit to the $\{\mathrm{CX}, U\}$ basis
(Table~\ref{tab:gate_resources}, Appendix~\ref{app:oracle_circuit}).  These values
are independent of the shot or query budget --- they characterise the
cost of a single circuit execution, not the number of executions --- and
are reported separately because implementation feasibility on
fault-tolerant hardware depends on gate count and depth in a way that
shot- or query-counting alone does not capture.  The model oracle's
$\mathcal{A}$ circuit is not included in this accounting. Its diagonal
phase-kickback construction decomposes into $O(2^n)$ elementary gates by
construction (Sec.~\ref{sec:oracle_design}) and is not intended as an
executable circuit.

\begin{figure*}[t]
  \centering
  \includegraphics[width=\textwidth]{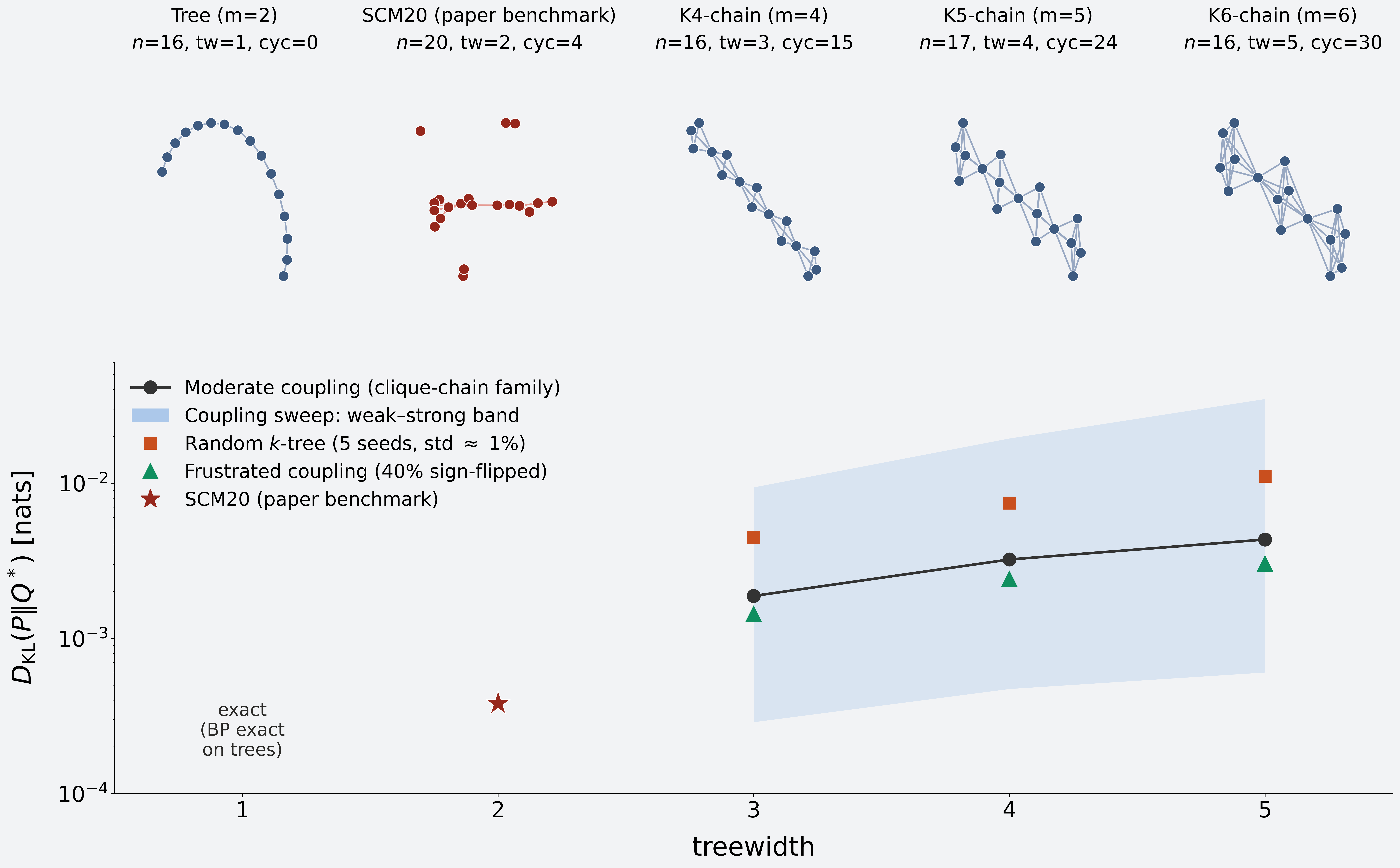}
  \caption{\label{fig:graph_family}%
    Robustness of \ac{bpacl} state preparation to graph topology and
    coupling structure. Top: force-directed layouts of the synthetic
    clique-chain family and \ac{scm20}, ordered by treewidth. Labels give
    the number of vertices $n$, treewidth (tw), and cycle rank (cyc).
    Bottom: $D_{\mathrm{KL}}(P\|Q^*)$ against treewidth, where lower values
    indicate a more accurate approximation. Dark-gray circles and the
    connecting line show the moderate-coupling clique-chain family, the red
    star shows \ac{scm20}, and the blue band
    shows the range obtained by sweeping the coupling magnitude from weak
    to strong. Orange squares report the mean over five random $k$-tree
    topologies, while green triangles show the
    frustrated case with $40\%$ of coupling signs flipped.}
\end{figure*}

\section{Results}
\label{sec:results}

We evaluate the full quantum pipeline on \ac{scm20} across four stages: state
preparation, oracle correctness, and tail-probability estimation at two
thresholds of increasing rarity.  All quantum circuits were implemented with
Qiskit 2.4.1~\cite{JavadiAbhari:2024} and simulated using Qiskit Aer 0.17.2's
AerSimulator~\cite{QiskitAer} in statevector mode (exact amplitudes); classical
baselines ran on the same host.  Exact ground truth is obtained by full
enumeration of all $2^{20} = 1\,048\,576$ Ising configurations.

\subsection{Robustness Across Graph Topologies}
\label{sec:state_prep_robustness}

Before examining \ac{scm20} in detail, we first test whether
\ac{bpacl}'s approximation quality generalises across graph topology,
rather than being specific to this one instance. \ac{scm20} has
treewidth exactly 2 (Sec.~\ref{sec:benchmark}); we evaluate the full
\ac{bpacl} pipeline --- \ac{bp} marginals, Chow--Liu tree, and
conditional-rotation circuit, unmodified from the construction in
Sec.~\ref{sec:state_prep} --- on a family of synthetic Ising graphs of
increasing treewidth, and find that state preparation stays accurate
across this range, not only for \ac{scm20}.

Each synthetic instance is a chain of complete-graph blocks of size $m$,
joined at single cut vertices (vertices whose removal disconnects the graph):
block size $m=2$ reduces to a tree,
and larger $m$ produces short, dense local cycles connected by narrow
bridges, the same qualitative structure independently found in
\ac{scm20} itself (Sec.~\ref{sec:model}). A clique $K_m$ (the complete
graph on $m$ vertices, where every pair of vertices is connected) has
treewidth exactly $m-1$, and joining cliques at single cut vertices does not
increase the treewidth beyond the largest block~\cite{KollerFriedman:2009}.
We additionally verify the treewidth numerically for every instance using
exact simplicial-vertex elimination, which successively removes vertices
whose neighbours form a clique. This procedure is exact here because all
instances, including \ac{scm20}, are chordal: every cycle of length at least
four has an edge joining two non-consecutive vertices. We cross-check the
results against independent treewidth upper-bound heuristics. For this
topology comparison, every instance
shares the same field and coupling magnitude, chosen to match the
\ac{scm20} calibration ($h_i \approx -3.5$, $J_{ij} \approx 0.45$), so
that only graph topology varies and the marginal failure probability
stays in the same rare-event range throughout. We relax this choice
separately below, varying coupling magnitude and sign instead of
topology. The upper panel of Fig.~\ref{fig:graph_family} shows the resulting
graph family.

The dark-gray clique-chain curve in the lower panel of
Fig.~\ref{fig:graph_family} shows that approximation error increases
monotonically with treewidth within this clique-chain family, spanning
more than three orders of magnitude in $D_{\mathrm{KL}}$. It is exact
to floating-point precision on a tree, as expected since \ac{bp} is
exact on trees, small but non-zero at \ac{scm20}'s treewidth of 2, and
grows steadily as denser local cycles are added. \ac{scm20} is not the
trivial tree case. It sits at a genuine, non-zero point on this trend,
which supports its use as the paper's primary
benchmark. Every point in this trend uses one topology per
treewidth level, with the same field and coupling magnitude and sign
throughout, so on its own this trend is evidence within the tested
family rather than a general law relating treewidth to \ac{bpacl}
accuracy.

We tested this further with three checks, summarized in the lower panel of
Fig.~\ref{fig:graph_family}, that vary topology, coupling magnitude, and
coupling sign independently. First, at treewidths 3--5, we replaced each
clique-chain topology with five random $k$-trees. The orange squares show the
mean $D_{\mathrm{KL}}$ over the five realizations. The mean divergence still
increases with treewidth and is roughly twice that of the corresponding
clique chain, indicating that the clique-chain family is a favourable rather
than typical case. Second, the blue band shows the effect of varying coupling
magnitude from weak to strong at fixed topology. This changes
$D_{\mathrm{KL}}$ by one to two orders of magnitude at the same treewidth, so
coupling strength is an independent factor within the cases tested. Third,
the green triangles show the frustrated cases, with $40\%$ of the coupling
signs flipped. Belief propagation converged in six to eight iterations in
every case, with accuracy comparable to the unfrustrated baseline. We did not
find evidence that sign frustration breaks the method at the coupling
strength and frustration fraction tested here.

This experiment tests only the classical state-preparation
approximation error in isolation. It does not re-run the full quantum
\ac{mlae} comparison of
Secs.~\ref{sec:results_tref}--\ref{sec:results_tstruct} on each
topology or coupling setting, which we leave to future work.

\subsection{State Preparation Accuracy}

\begin{figure*}[t]
  \centering
  \includegraphics[width=0.80\textwidth]{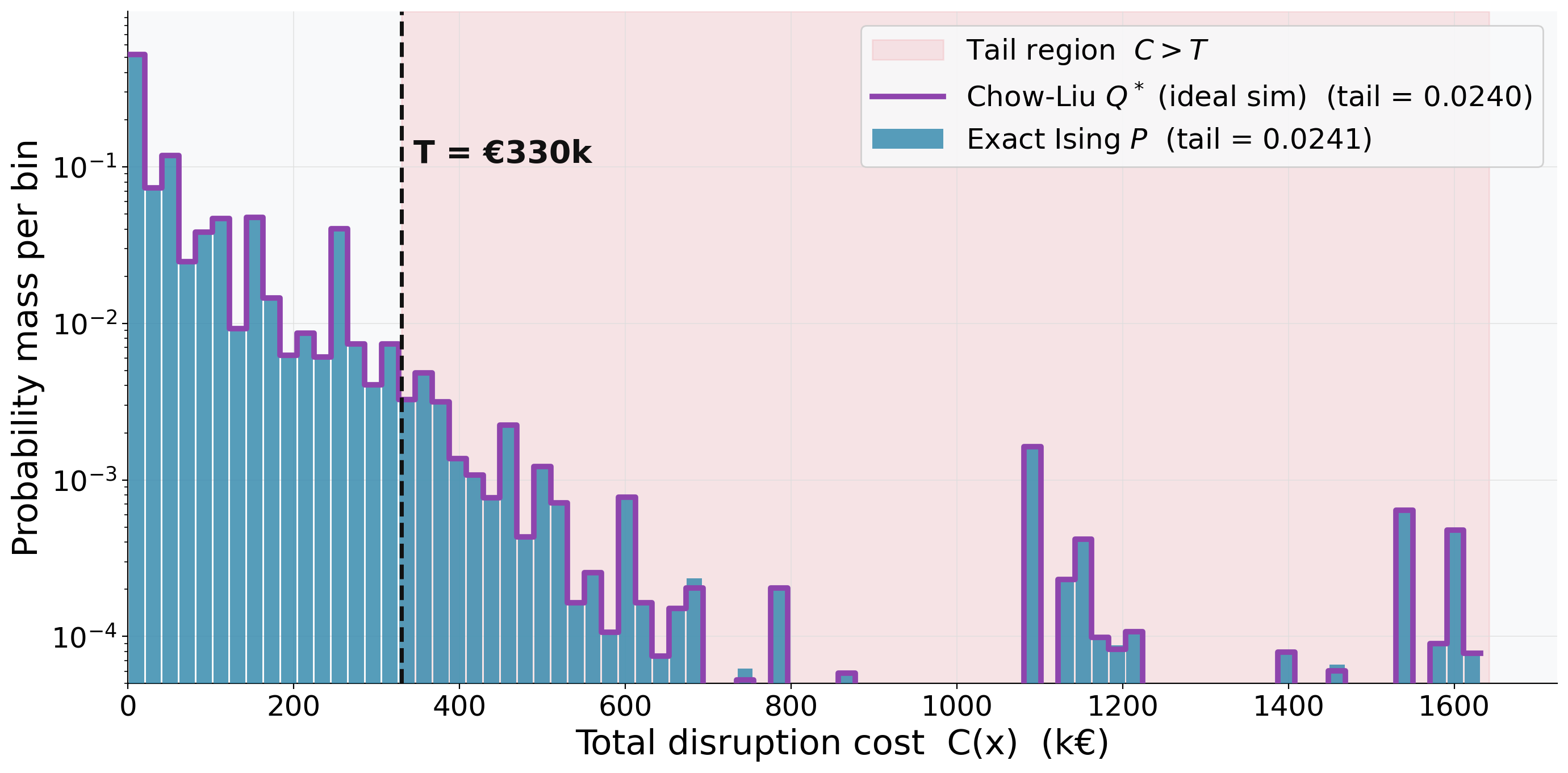}
  \caption{\label{fig:cost_distribution}%
    Probability-mass distribution of the disruption cost over all
    configurations of \ac{scm20}, shown on a logarithmic scale.
    Blue bars: exact Ising distribution $P$.
    Purple step line: Chow-Liu approximation $Q^*$ encoded in the
    state-preparation circuit.
    The dashed vertical line marks the reference threshold $T_\mathrm{ref}$;
    the shaded region to its right marks the rare tail targeted by \ac{mlae}.
    The two distributions are nearly indistinguishable on this scale.}
\end{figure*}

Having shown this trend holds broadly, we now look closely at
\ac{scm20} itself, the primary benchmark used throughout the rest of
this paper, and find that the prepared state matches the exact
distribution almost exactly. For the present model,
belief propagation converges in 7 iterations and runs in under 1\,ms. Of
the model's 20 original Ising edges, the Chow-Liu maximum-spanning-tree
algorithm retains 16 (all edges of the two two-node components, plus 14
of the 15-node core component's 18 edges) and drops the remaining 4,
which carry mutual information below $2 \times 10^{-4}$ nats. Because
the graph has four disconnected components (Sec.~\ref{sec:benchmark}),
a single 19-edge tree spanning all 20 variables additionally requires 3
new edges, absent from the original Ising graph, connecting the four
components; cross-component pairs are exactly independent under the
Ising factorisation, so these added edges carry zero true mutual
information. The resulting 20-qubit circuit
has depth 43 and 38~CX gates.

Full enumeration confirms that the Chow--Liu approximation $Q^*$ closely
matches the exact Ising distribution $P$, with
$\KL{P}{Q^*}=3.82\times10^{-4}$ nats. The maximum absolute error in the single-node marginals
obtained by \ac{bp}, relative to exact enumeration, is
below $2 \times 10^{-5}$. Thus, \ac{bp} has effectively
converged to the exact marginals on this graph. At
the reference threshold, the state-preparation-bias term of the
error decomposition (Eq.~\eqref{eq:error_decomposition}) evaluates to
$-3.0 \times 10^{-5}$, corresponding to a relative underestimate of
$0.12\%$.
This bias is primarily caused by the four omitted edges rather
than by inaccuracies in \ac{bp}, and it remains far smaller than
the statistical uncertainty of a finite-sample Monte Carlo
estimator at the budgets considered here.

Figure~\ref{fig:cost_distribution} visualises the full disruption-cost
spectrum on a logarithmic scale, making the multi-decade tail structure
accessible.  Several features of the discrete cost model are directly
visible: the 20 binary failure variables produce only a finite set of
reachable cost sums, leaving genuine empty bands near
$\text{EUR}\,800\text{k}$ and $\text{EUR}\,1.3\text{M}$ that are not
binning artefacts.  The log scale also confirms that $Q^*$ (purple step
line) tracks $P$ (teal bars) across three orders of magnitude, from the
bulk near zero to the tail.

\subsection{Oracle Verification}
\label{sec:oracle_verification}

Table~\ref{tab:oracle_comparison} summarises the two oracle
implementations.  For each oracle we read the ancilla amplitude $a =
P(\text{ancilla}=1)$ directly from the statevector before any \ac{mlae}
sampling; this isolates oracle correctness from estimator variance.

\begin{table}[h]
\caption{\label{tab:oracle_comparison}%
  Oracle properties on \ac{scm20}.  The model oracle (DiagonalGate
  phase-kickback) targets $T_\mathrm{ref}$ and encodes the tail
  probability directly from the cost lookup table.  The structural oracle
  (Toffoli-based) targets $T_\mathrm{struct}$ and marks configurations
  by explicit gate-level logic, without a classical bypass.
  $a$ is the amplitude read from the statevector;
  $P(C>T)$ is the exact Ising ground truth. The model oracle's
  DiagonalGate is never decomposed in simulation, so no real depth or
  gate count is measured for it.}
\begin{tabular}{lcc}
\toprule
                                   & Model oracle            & Structural oracle       \\
\midrule
Threshold $T$                      & EUR 330\,000            & EUR 779\,999            \\
Exact $P(C>T)$                     & 0.024070                & 0.005229                \\
Oracle gate type                   & DiagonalGate            & Toffoli                 \\
Qubits                             & 21                      & 28                      \\
Oracle ($F_T$) depth               & $O(2^n)$                & 36                      \\
$A = F_T{\cdot}(R{\otimes}I)$ depth  & $O(2^n)$   & 75                      \\
CX gates in $A$                    & $O(2^n)$                & 52                      \\
CCX / MCX gates in $A$             & n/a                     & 17 / 12                 \\
Simulated amplitude $a$            & 0.024040                & 0.005076                \\
$|a - Q^*(C>T)|$                   & $1.0 \times 10^{-17}$   & $1.5 \times 10^{-4}$    \\
$|a - P(C>T)| / P(C>T)$           & 0.12\%                  & 2.95\%                  \\
\bottomrule
\end{tabular}
\end{table}

For the model oracle the simulated amplitude $a = 0.024040$ agrees with
the Chow-Liu reference $Q^*(C > T_\mathrm{ref}) = 0.024040$ to
$|a - Q^*| = 10^{-17}$, i.e.\ floating-point machine precision.
Because the DiagonalGate oracle is constructed from the complete classical
cost table for all $2^{20}$ configurations, this agreement is expected:
the oracle exactly marks every state with $C(x) > T_\mathrm{ref}$, so
the only source of error relative to the exact Ising ground truth $P(C>T) =
0.024070$ is the state-preparation-bias term of
Eq.~\eqref{eq:error_decomposition},
which arises from the four Chow-Liu dropped edges.
The oracle itself contributes zero additional error; any non-zero oracle
error would manifest as $|a - Q^*| \gg 0$, which is not observed.

The model oracle's DiagonalGate is never decomposed in simulation, so
Table~\ref{tab:oracle_comparison} reports only its asymptotic $O(2^n)$
implementation cost, not measured numerical depth or gate counts. The model
oracle therefore serves only as a classical validation reference. The
structural oracle is the only path with a real, transpiled gate count. Its cost
grows additively with Grover
depth $k$ and reaches 2\,323~CX gates at $k=3$. This places any
multi-iteration \ac{mlae} schedule in the fault-tolerant regime for
practical hardware execution. Implementation feasibility is determined
by these gate-level costs rather than by the shot or query budget. The
full breakdown across $k$ is given in
Appendix~\ref{app:oracle_circuit} (Table~\ref{tab:gate_resources}).

For the structural oracle the amplitude $a = 0.005076$ sits $2.9\%$ below
the exact ground truth $P(C>T) = 0.005229$. This is the under-marking
offset $\epsilon_\mathrm{miss}$ defined in Sec.~\ref{sec:oracle_struct}.
The Chow-Liu reference $Q^*(C > T_\mathrm{struct}) = 0.005224$ confirms
that the overwhelming majority of the gap is oracle under-marking
rather than state-preparation bias. This floor is constant across
\ac{mlae} repetitions and distinguishable from statistical variance, as
demonstrated in Sec.~\ref{sec:results_tstruct}.

\subsection{Tail-Probability Estimation at \texorpdfstring{$T_\mathrm{ref}$}{T\_ref}}
\label{sec:results_tref}

\begin{SCfigure}[24][t]
  \centering
  \includegraphics[width=0.50\columnwidth]{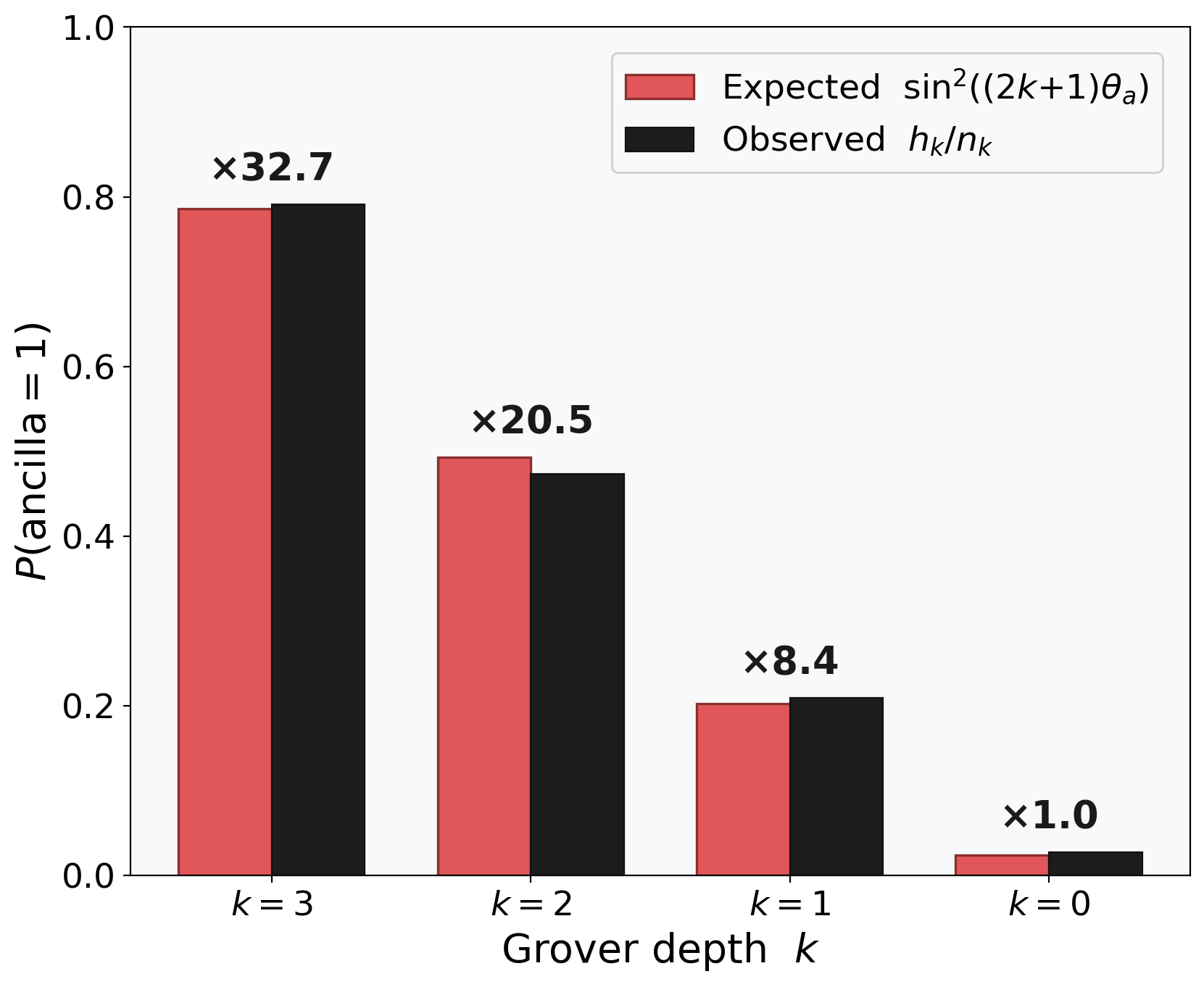}
  \caption{\label{fig:mlae_amplification}
    Grover amplification in the \ac{mlae} circuit for the model oracle,
    displayed in decreasing-amplification order ($k=3$ to $k=0$, left
    to right).  Red bars show the theoretical prediction; dark bars
    show the observed $\ket{1}$ ancilla fraction from simulated
    shots.  Annotated factors above each bar pair confirm that
    $\mathcal{Q}^k\mathcal{A}$ correctly implements $k$ Grover
    iterations.}
\end{SCfigure}

Before reporting estimation accuracy we verify that the Grover rotation
underlying \ac{mlae} is correctly assembled in simulation.
Figure~\ref{fig:mlae_amplification} shows $P(\text{ancilla}=1)$ for
the model oracle across all four
Grover depths.  Observed ancilla fractions (black bars, 2\,000 shots each)
agree with the theoretical prediction (red bars) within $\pm 2\sigma$
binomial uncertainty at every depth, confirming that the composed circuit
$\mathcal{Q}^k\mathcal{A}$ correctly implements $k$ applications of the Grover operator.
The amplification factors $\times 8.4$, $\times 20.5$, and $\times 32.7$
at $k=1,2,3$ reflect the increased per-shot Fisher information at each
Grover depth (Eq.~\eqref{eq:fisher_k}, Sec.~\ref{sec:results_schedule_scaling}).
Each additional application of $\mathcal{Q}$ rotates the ancilla amplitude
through a larger angle for the same underlying $\theta_a$, so a single
shot at depth $k$ carries $(2k+1)^2$ times more information about
$\theta_a$ than a depth-$0$ shot.  Without this amplification the estimator
would be no more informative than ordinary Monte Carlo at the same shot
count. Sec.~\ref{sec:results_schedule_scaling} shows precisely the
relationship between this per-shot gain and the asymptotic complexity
under the fixed schedule used throughout this paper.

\begin{figure*}[t]
  \centering
  \begin{subfigure}[t]{0.48\textwidth}
    \centering
    \includegraphics[width=\textwidth]{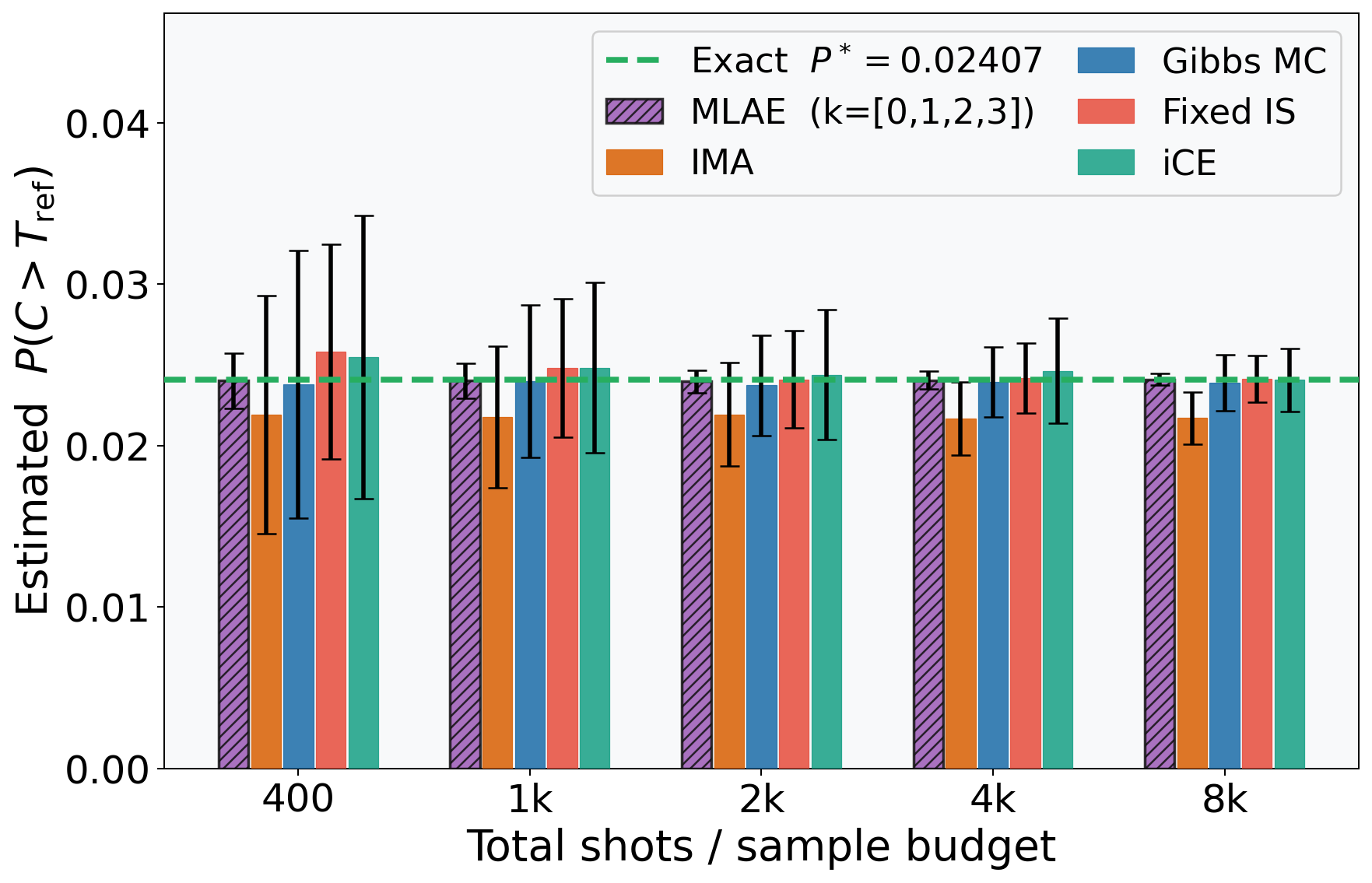}
    \caption{\label{fig:comparison_tref} Model oracle, reference threshold $T_\mathrm{ref}$.}
  \end{subfigure}
  \vspace{1.2em}
  \begin{subfigure}[t]{0.48\textwidth}
    \centering
    \includegraphics[width=\textwidth]{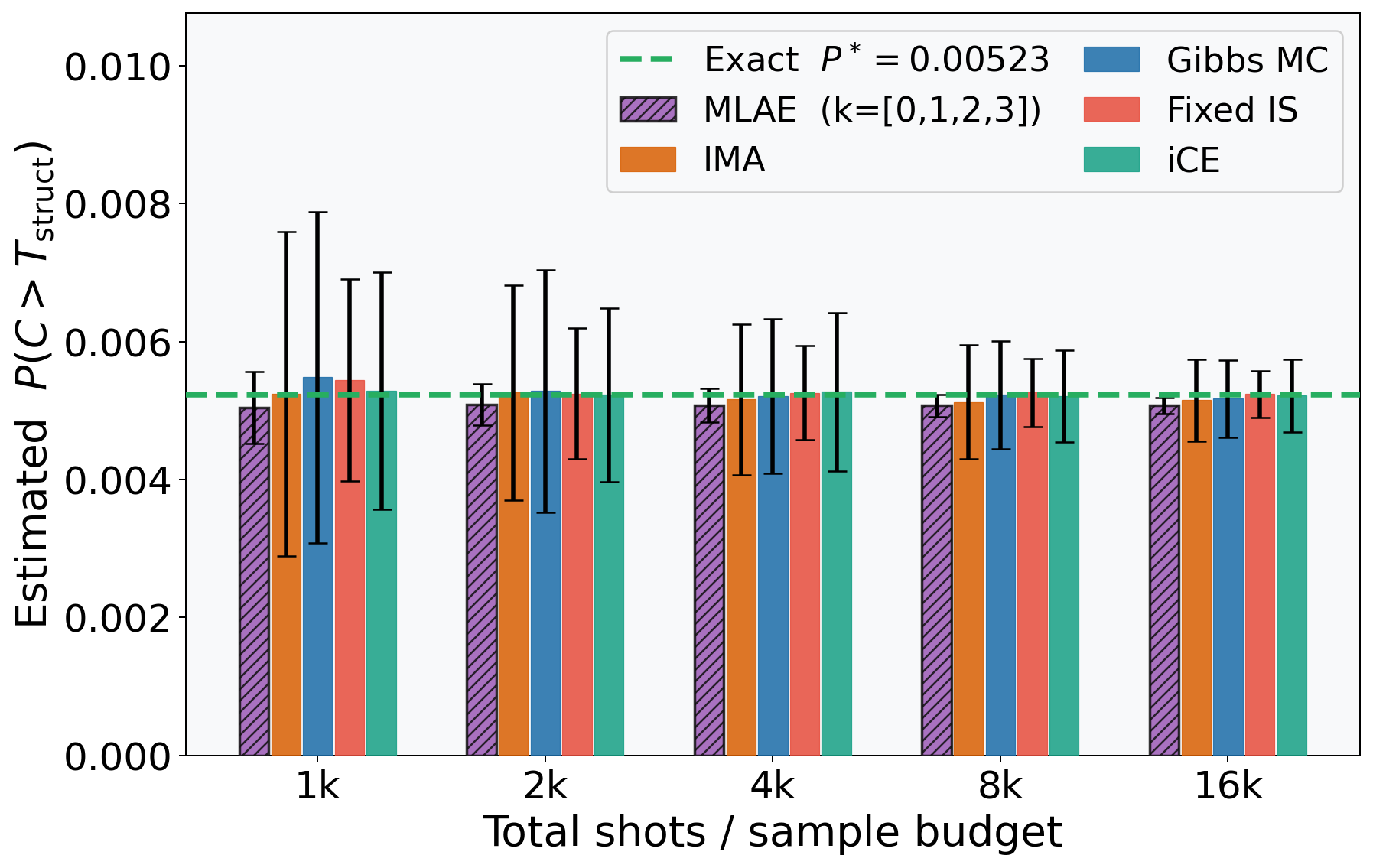}
    \caption{\label{fig:comparison_tstruct} Structural oracle, rarer threshold $T_\mathrm{struct}$.}
  \end{subfigure}
  \caption{\label{fig:comparison}%
    Tail-probability estimates $\hat{p}$ as a function of sample/shot budget
    $N$; error bars show one standard deviation over independent repetitions.
    The dashed green line marks the exact ground truth in each panel.}
\end{figure*}

Figure~\ref{fig:comparison_tref} shows the estimated tail probability
$\hat{p}$ obtained using five different
methods and five circuit-shot or sample budgets. For each
combination of method and budget, the experiment is repeated independently
200 times. These repetitions provide a sufficiently large ensemble for
assessing both the central tendency and the run-to-run variability of the
resulting probability estimates. The exact ground-truth value
$P(C>T_\mathrm{ref})$ is shown alongside the estimates to facilitate a
direct comparison of the accuracy, bias, and convergence behaviour of the
different methods as the available budget increases.
For \ac{mlae}, the amplification schedule is
$k\in\{0,1,2,3\}$. The total number of shots $N$ is divided equally
among the four amplification depths, so that each depth receives the same
number of shots. This allocation is used consistently for all values of
$N$, allowing the performance of \ac{mlae} to be compared across budgets
under a fixed scheduling and resource-allocation strategy.

At the largest budget in our sweep ($N = 8\,\mathrm{k}$), \ac{mlae} settles
to within $0.1\%$ of the exact ground truth with the smallest standard
deviation among all five methods ($\sigma = 0.00037$).
The same advantage is already visible at the smallest budget ($N = 400$),
where \ac{mlae}'s standard deviation is roughly four times lower than that
of the best-variance classical competitor at that scale.
Within statistical noise, this ratio remains constant across the full range
of $N$: for $N=400$, $1,\mathrm{k}$, $2,\mathrm{k}$, $4,\mathrm{k}$,
and $8,\mathrm{k}$, the corresponding values
are $3.9\times$, $4.0\times$, $4.3\times$, $3.9\times$,
and $3.9\times$, respectively. Thus, the ratio neither narrows
nor widens as $N$ increases.
This is exactly what Sec.~\ref{sec:results_schedule_scaling}'s
Fisher-information argument predicts for a \emph{fixed} Grover-depth
schedule: both \ac{mlae} and every classical estimator here scale as
$O(N^{-1/2})$, so their ratio is asymptotically flat.  We do not observe or
claim the gap narrowing toward the ideal quadratic rate; that would
require the schedule itself to deepen with the budget, which we address
separately in Sec.~\ref{sec:results_schedule_scaling}.

\ac{naivemc} exhibits a persistent negative bias of approximately
$-9\%$ to $-10\%$ that does not shrink with budget. This bias is
structural because \ac{naivemc} draws $x_i\sim
\mathrm{Bernoulli}(q_i)$ independently from the single-node marginals.
It therefore underrepresents positively correlated simultaneous
failures and depresses $\hat{P}(C>T_\mathrm{ref})$ by a fixed fraction
regardless of $N$. \ac{gibbsmc} avoids this bias by sampling from the
full joint distribution and stays within $1$--$2\%$ of the exact value
at every budget tested. At $N=8\,\mathrm{k}$, its relative error is
$-0.8\%$, although its standard deviation of $\sigma=0.00175$ remains
roughly five times larger than that of \ac{mlae}.

At large $N$, \ac{fixedis} achieves the tightest variance among the
classical methods. At $N=8\,\mathrm{k}$, it has
$\sigma=0.00145$ and is essentially unbiased at $+0.2\%$. Its
performance relative to \ac{mlae} deteriorates at smaller budgets. At
$N=400$, \ac{fixedis} reaches $\sigma=0.00666$, compared with
$\sigma=0.00170$ for \ac{mlae}. The fixed proposal provides little
variance reduction when few samples are available to cover the tail.
\ac{ice} also performs poorly at small budgets because it requires
several warm-up rounds to adapt its proposal. At $N=400$, it has the
highest variance of the four classical methods, with
$\sigma=0.00879$, roughly five times that of \ac{mlae}. Its residual
offset has largely disappeared by $N=8\,\mathrm{k}$ as the
cross-entropy adaptation converges.

\subsection{Tail-Probability Estimation at \texorpdfstring{$T_\mathrm{struct}$}{T\_struct}}
\label{sec:results_tstruct}

Figure~\ref{fig:comparison_tstruct} repeats the comparison at the rarer
threshold $T_\mathrm{struct} = \text{EUR}\,779\,999$ (exact ground truth
$P(C>T_\mathrm{struct}) = 0.005229$, roughly one-fifth the frequency of
$T_\mathrm{ref}$),
using the structural oracle.  The rarer tail amplifies the difficulty for
classical methods but also reveals the systematic under-marking floor
identified in Sec.~\ref{sec:oracle_verification}.

The effect of rarity is already apparent at the smallest budget,
$N=1\,\mathrm{k}$. With $P(C>T_\mathrm{struct}) \approx 0.52\%$, a budget
of one thousand samples produces only about five tail events on average. Consequently,
every classical method shows visibly elevated variance relative to its
$T_\mathrm{ref}$ result at the same budget, with $\sigma$ between
$0.0015$ and $0.0024$. With $R=200$ repetitions, the relative errors of
the ensemble means with respect to the exact ground truth are all below
$5\%$ in magnitude: \ac{naivemc} $+0.2\%$, \ac{gibbsmc} $+4.8\%$,
\ac{fixedis} $+4.0\%$, and \ac{ice} $+1.1\%$.

At the same budget, \ac{mlae} achieves $\sigma=0.00052$ and a relative
error of $-3.6\%$. It is roughly $4.5\times$ more precise than
\ac{naivemc} or \ac{gibbsmc} because Grover amplification encodes the
tail probability in the rotation angle of a quantum state rather than
counting rare events directly.

As budgets grow, \ac{mlae} converges to a deterministic floor of
$-3.0\%$ relative error at $N=16\,\mathrm{k}$, with
$\sigma=0.00012$. This closely matches the combined contribution of
oracle under-marking and state-preparation bias
identified in Table~\ref{tab:oracle_comparison}. The floor is reproducible across repetitions and
therefore distinguishable from statistical noise. It could in principle
be removed by a one-time calibration against a small exact benchmark.
We leave this calibration to future work and report the floor as a known
limitation of the structural oracle at its current level of completeness.

Despite the floor, \ac{mlae}'s variance advantage remains clear
throughout the budget range. It is approximately $3$--$5\times$ relative
to the best classical competitor, which is \ac{fixedis} at large $N$.
This repeats the constant-factor pattern observed at $T_\mathrm{ref}$
and follows from the same fixed-schedule mechanism
(Sec.~\ref{sec:results_schedule_scaling}). \ac{gibbsmc} and
\ac{naivemc} provide only modest improvements over one another at this
threshold because direct-counting methods struggle to estimate a tail
probability below $1\%$. \ac{fixedis} performs better by concentrating
samples in the tail through reweighting. A threshold-specific recalibration
toward a higher expected failure count was counterproductive, reducing the
effective sample size by approximately half and increasing RMSE by
$30$--$45\%$ across all tested budgets. Overall, \ac{mlae} maintains the
lowest standard deviation of all five methods at every budget tested.

\subsection{Query-Normalised Comparison}
\label{sec:results_query}

\begin{figure*}[t]
  \centering
  \begin{subfigure}[t]{0.48\textwidth}
    \centering
    \includegraphics[width=\textwidth]{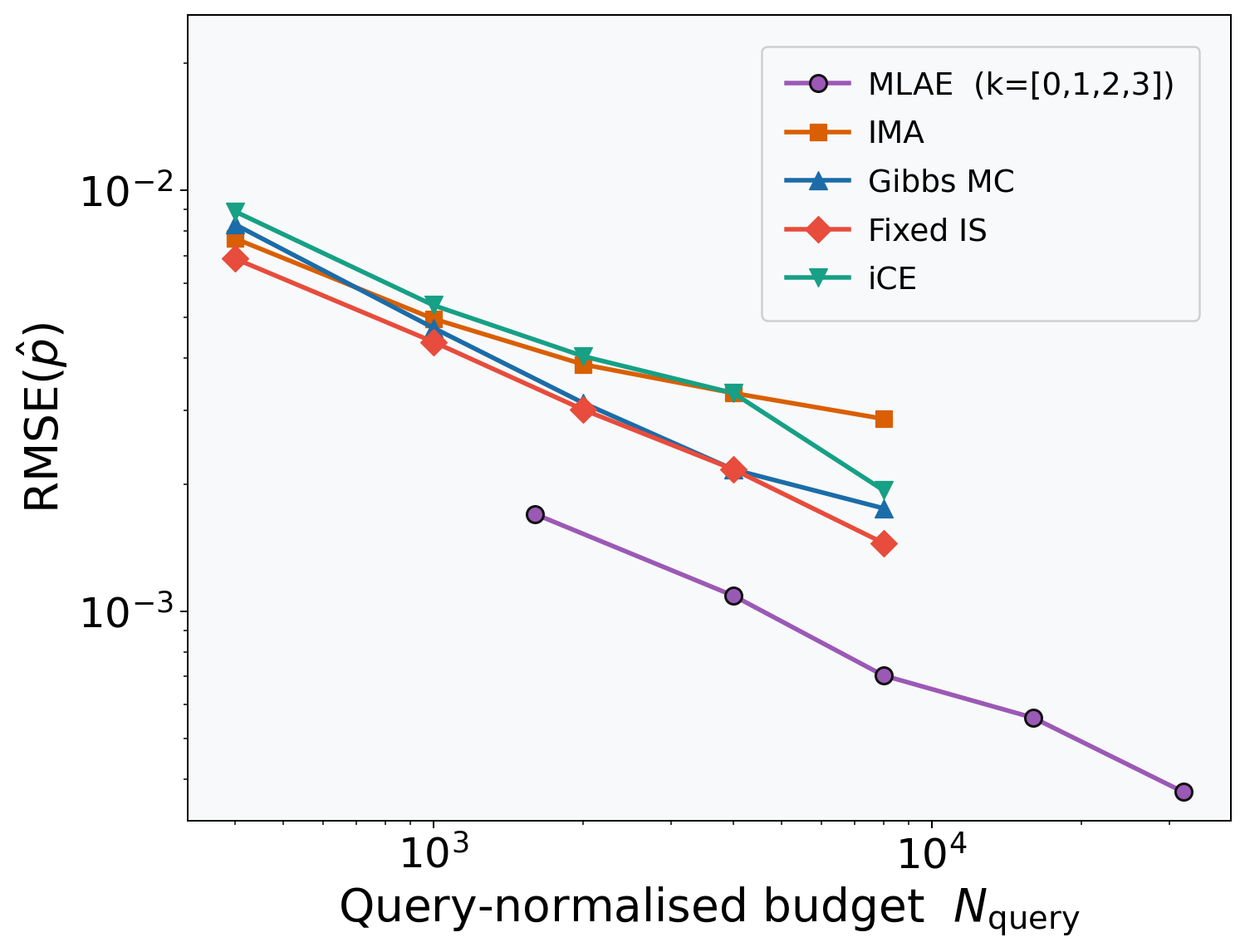}
    \caption{\label{fig:rmse_query_tref} Model oracle, reference threshold $T_\mathrm{ref}$.}
  \end{subfigure}
  \hfill
  \begin{subfigure}[t]{0.48\textwidth}
    \centering
    \includegraphics[width=\textwidth]{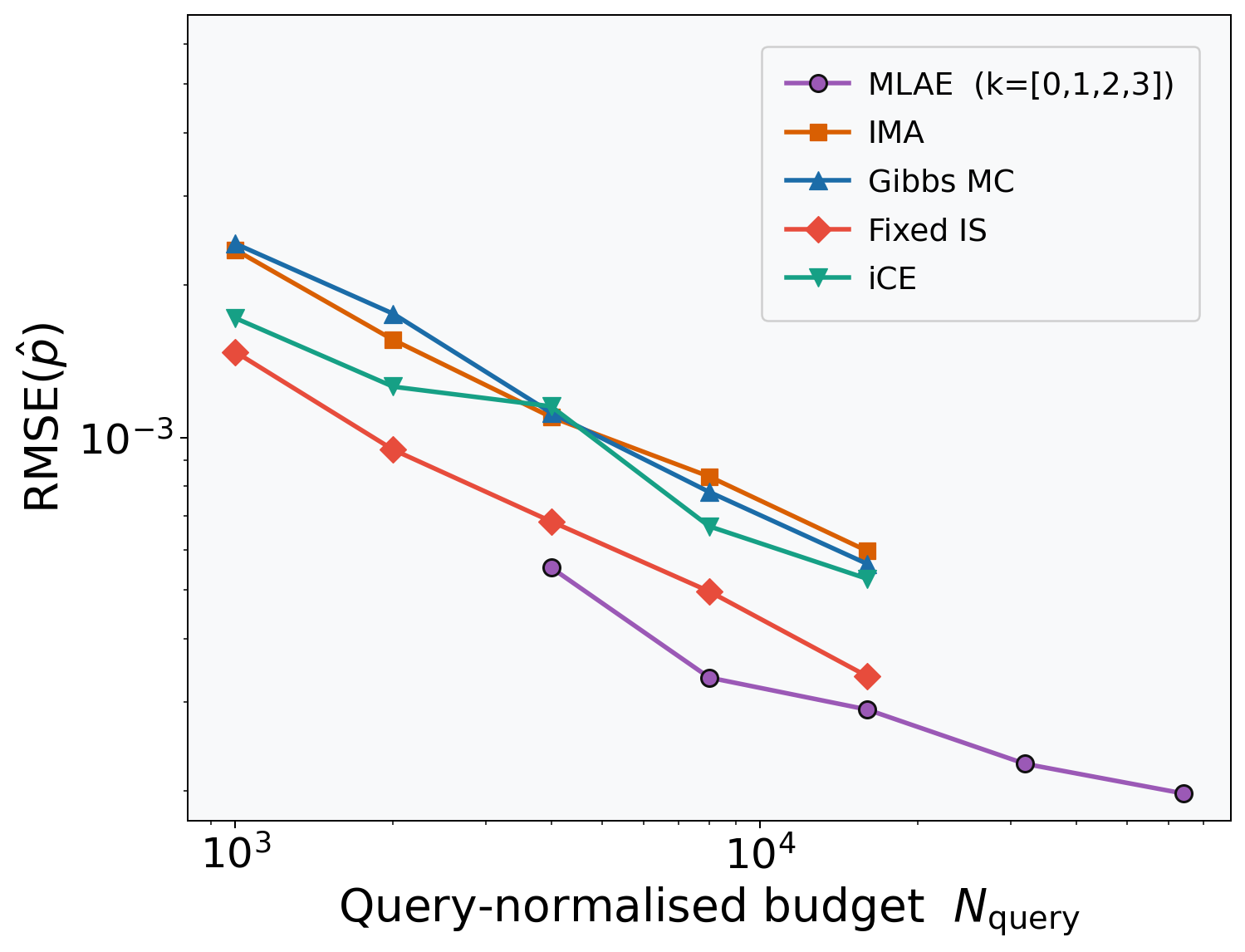}
    \caption{\label{fig:rmse_query_tstruct} Structural oracle, rarer threshold $T_\mathrm{struct}$.}
  \end{subfigure}
  \caption{\label{fig:rmse_query}%
    RMSE against the query-normalised budget $N_\mathrm{query}$.
    Lines connect swept budget points; least-squares slopes of
    $\log_{10}\mathrm{RMSE}$ against $\log_{10}N_\mathrm{query}$ over the
    shown range are reported in the text below.  A slope of $-1$ would
    indicate the ideal $\mathrm{RMSE} = O(N_\mathrm{query}^{-1})$ rate;
    $-1/2$ indicates ordinary classical $O(N^{-1/2})$ statistical
    scaling.}
\end{figure*}

The comparisons above are shot-for-sample: every method is credited with
one unit of budget per completed circuit execution or classical sample.
As discussed in Sec.~\ref{sec:budget_query}, this is not a fair
query-complexity comparison, because a shot at Grover depth $k$ consumes
$2k+1$ calls to $\mathcal{A}$ rather than one.  Figure~\ref{fig:rmse_query}
replots both benchmarks against the query-normalised budget
$N_\mathrm{query}$ of Eq.~\eqref{eq:n_query}, using the root-mean-square
error
\begin{equation}
  \mathrm{RMSE}(\hat{p}) = \sqrt{\mathrm{Bias}^2 + \mathrm{Var}},
  \label{eq:rmse}
\end{equation}
with
\begin{equation}
  \mathrm{Bias} = \overline{\hat p} - p_\mathrm{tail},
  \qquad
  \mathrm{Var} = \frac{1}{M}\sum_{m=1}^{M}\bigl(\hat p^{(m)} - \overline{\hat p}\bigr)^2,
  \label{eq:bias_var}
\end{equation}
where $\hat p^{(m)}$ is the estimate from the $m$-th of $M$ independent
repetitions, $\overline{\hat p} = \tfrac{1}{M}\sum_{m=1}^{M}\hat p^{(m)}$
is their sample mean, and $p_\mathrm{tail}$ is the exact ground truth
(Eq.~\eqref{eq:ptail}), rather than the raw estimate, since \ac{mlae}'s
deterministic floors make bias and variance separately relevant.  For the fixed schedule
$\mathcal{K}=\{0,1,2,3\}$ used throughout, $N_\mathrm{query} =
4N_\mathrm{shots}$ (Eq.~\eqref{eq:n_query_factor}), so every \ac{mlae}
point in Fig.~\ref{fig:rmse_query} is shifted a factor of four to the
right of its position in Figs.~\ref{fig:comparison_tref}--\ref{fig:comparison_tstruct}.

At $T_\mathrm{ref}$, the fitted \ac{mlae} slope is $-0.50$, which lies
\emph{inside} the range spanned by the four classical methods ($-0.33$ to
$-0.53$) rather than near the ideal $-1$.  This confirms that the fixed,
bounded-depth schedule used in this paper does not exhibit the asymptotic
$O(N_\mathrm{query}^{-1})$ rate associated with amplitude estimation's
quadratic query advantage. Measured this way, \ac{mlae}'s empirical
convergence is statistically indistinguishable from ordinary
$O(N^{-1/2})$ behaviour.  What does survive query-normalisation is a
consistent constant-factor improvement: at the two budgets where an
\ac{mlae} point and a classical sweep point coincide exactly
($N_\mathrm{query}=4\,000$ and $8\,000$), \ac{mlae}'s RMSE is
$2.0\times$ and $2.1\times$ lower, respectively, than the best-performing
classical method at that same query count (\ac{gibbsmc} at $4\,000$,
\ac{fixedis} at $8\,000$).
Therefore, Fig.~\ref{fig:comparison_tref} demonstrates a genuine but finite-budget
advantage, unrelated to any asymptotic improvement in complexity.

At $T_\mathrm{struct}$, the \ac{mlae} slope flattens further to around $-0.35$, compared
with classical slopes ranging from $-0.43$ to $-0.54$. This slower decrease reflects
the onset of the deterministic under-marking floor
identified in Sec.~\ref{sec:results_tstruct}. As
the statistical component shrinks, the bias floor increasingly limits further
reductions in RMSE.
The advantage of \ac{mlae} over the best classical method—\ac{fixedis} at every
query count considered—is therefore modest and non-monotonic. The corresponding
improvement factors are $1.2\times$, $1.5\times$, and $1.2\times$ at
$N_\mathrm{query}=4\,000$, $8\,000$, and $16\,000$, respectively. This behaviour
is consistent with both estimators approaching their characteristic error
floors over this range: \ac{mlae} is limited by deterministic bias, whereas
the variance of \ac{fixedis} has fallen to a comparable level.
At $N_\mathrm{query}=16\,000$, for example, the
RMSE is $2.9\times10^{-4}$ for \ac{mlae} and $3.4\times10^{-4}$ for \ac{fixedis}.

More generally, this result illustrates the design principle discussed
qualitatively in Sec.~\ref{sec:discussion}: the accuracy of state preparation and
the event oracle limits the useful Grover depth, and hence the useful
query budget. The oracle and state-preparation biases combine
algebraically into a single net deterministic bias
(Eq.~\eqref{eq:error_decomposition}), which together with the
statistical variance determines the total RMSE (Eq.~\eqref{eq:rmse}).
Once that net bias exceeds the shrinking statistical variance, reducing
the statistical component further yields little improvement in RMSE.

\subsection{Does a Deeper Fixed Schedule Approach the Ideal Rate?}
\label{sec:results_schedule_scaling}

Section~\ref{sec:results_query} shows that the fixed schedule
$\mathcal{K}=\{0,1,2,3\}$ does not exhibit the ideal
$\mathrm{RMSE}=O(N_\mathrm{query}^{-1})$ rate.  A natural question is
whether this is simply because $\mathcal{K}$ is too shallow. Perhaps a
fixed schedule with larger $k_\mathrm{max}$ would recover the ideal rate.
We test this directly by repeating the model-oracle sweep of
Sec.~\ref{sec:results_tref} for three fixed schedules of increasing
maximum depth, $\mathcal{K}_A=\{0,1\}$, $\mathcal{K}_B=\{0,1,2,3\}$, and
$\mathcal{K}_C=\{0,1,2,4,8\}$, each with 50 repetitions per budget point.

The answer is structural, not a matter of insufficient depth.  A
Fisher-information argument, given in full in
Appendix~\ref{app:schedule_scaling}, shows that for any \emph{fixed}
Grover-depth schedule the total information about $\theta_a$ grows only
linearly with the query budget, so the resulting standard error can
never do better than the classical rate,
\begin{equation}
  \mathrm{RMSE}(\hat a) = \Theta\bigl(N_\mathrm{query}^{-1/2}\bigr).
  \label{eq:rmse_fixed_schedule}
\end{equation}
This holds regardless of how large $k_\mathrm{max}$ is, as long as the
set of depths used stays fixed while the shot budget grows. A deeper
fixed schedule changes only the constant prefactor in this rate, not
its exponent, and the same conclusion holds for any fixed allocation of
shots across depths, not only the equal split used here.

\begin{SCfigure}[24][t]
  \centering
  \includegraphics[width=0.5\columnwidth]{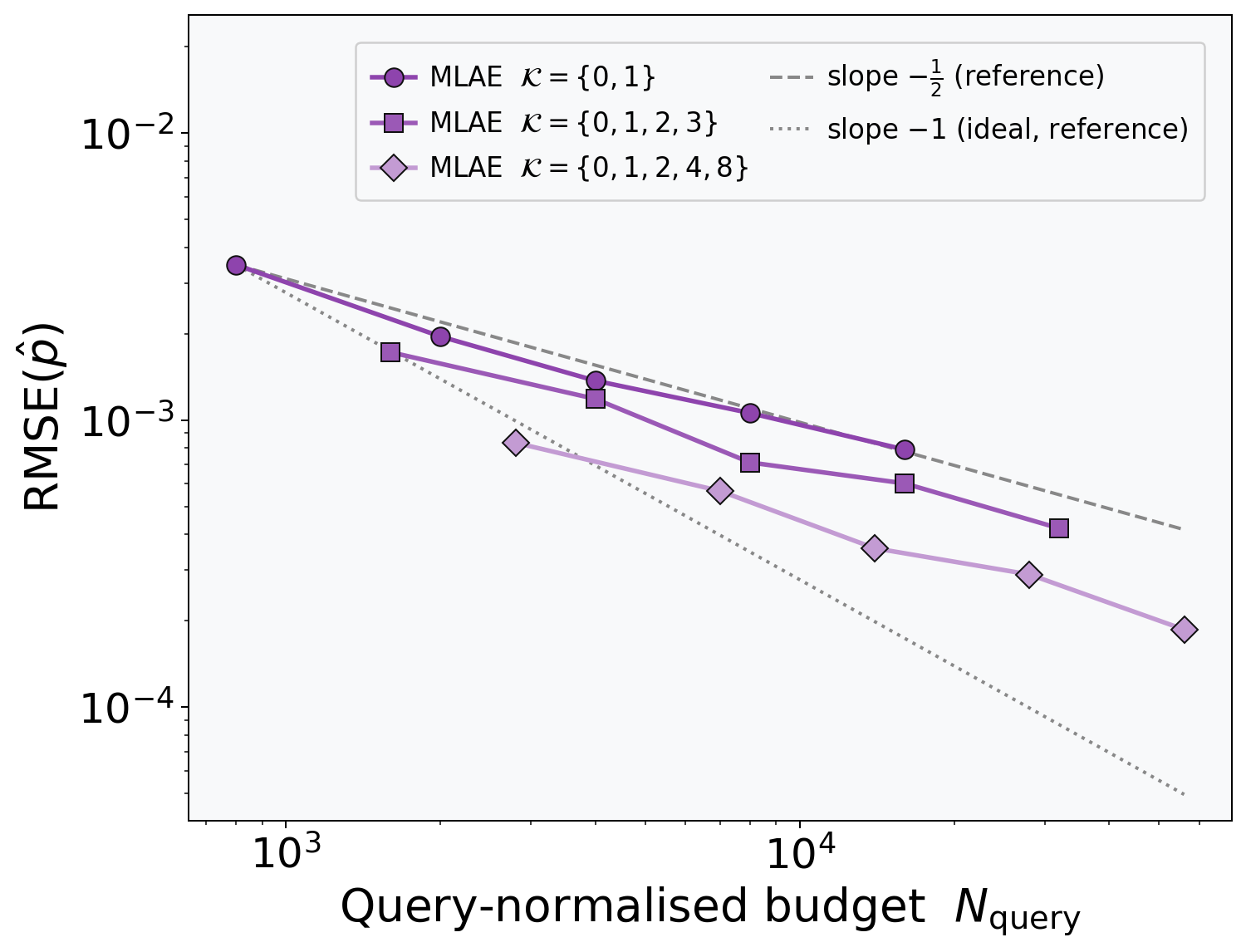}
  \caption{\label{fig:schedule_scaling}%
    RMSE against $N_\mathrm{query}$ for three fixed schedules of
    increasing maximum depth ($k_\mathrm{max}=1,3,8$), model oracle,
    $T_\mathrm{ref}$.  Dashed and dotted grey lines are slope $-1/2$ and
    $-1$ references anchored at $\mathcal{K}_A$'s first point.  All three
    fitted slopes track the $-1/2$ reference; none approaches $-1$.}
\end{SCfigure}

Figure~\ref{fig:schedule_scaling} confirms this empirically. The fitted
slopes are $-0.488$ for $\mathcal{K}_A$, $-0.475$ for
$\mathcal{K}_B$, and $-0.496$ for $\mathcal{K}_C$. These values are
statistically indistinguishable from one another and from $-1/2$, even
though $k_\mathrm{max}$ increases eightfold from $\mathcal{K}_A$ to
$\mathcal{K}_C$.

Increasing the maximum depth reduces the RMSE prefactor but leaves
the asymptotic exponent unchanged. At a
matched $N_\mathrm{query}$, the RMSE of $\mathcal{K}_C$ remains below
those of $\mathcal{K}_A$ and $\mathcal{K}_B$ throughout the swept
range. This confirms that the finite-budget constant-factor improvement
identified in Sec.~\ref{sec:results_query} persists as depth increases
without becoming a change in asymptotic rate.

The same Fisher-information argument
(Appendix~\ref{app:schedule_scaling}) also identifies what is required
to recover the ideal rate. The maximum depth $k_\mathrm{max}$ must grow
with the query budget. For example, it could double at each successive
budget point through the schedule
$\mathcal{K}_m=\{0,1,2,4,\ldots,2^{m-2}\}$ as $m$ increases. This differs
from allocating additional shots to a fixed schedule, however deep that
schedule may be. We do not pursue the growing-depth regime because its
circuit costs would be far beyond hardware-relevant depths and it would
require a separate analysis of \ac{mlae} branch-aliasing errors.

The schedules tested here span the depths for which this paper's
structural oracle and state-preparation circuits are themselves already
characterised (Sec.~\ref{sec:oracle_design},
Table~\ref{tab:gate_resources}), and are sufficient to establish the
central point of this section: the constant-factor advantage reported
throughout this paper is real and persists with depth, but is
mechanistically distinct from, and should not be conflated with, the
asymptotic quadratic query advantage.

\section{Discussion}
\label{sec:discussion}

\subsection{Scope of the Empirical Comparison}

The results of Sec.~\ref{sec:results} show that \ac{mlae} achieves
lower estimation error than all four classical Monte Carlo baselines
in the medium-to-large budget regime.  The scope of this acceleration
claim, however, must be stated precisely.  The comparison made in this
paper is against classical \emph{sampling-based} estimators, whose
statistical error decreases as $O(N^{-1/2})$ in the number of oracle
queries or samples $N$.  It is not a claim of advantage over every
possible classical algorithm on the particular $n=20$ benchmark
instance.

This distinction matters because the benchmark graph, although
disconnected and containing four independent cycles (cyclomatic
number 4), has treewidth exactly $\mathrm{tw}=2$. Its cycles are short
and confined to small blocks rather than forming one large loop.
Consequently, variable-elimination or junction-tree methods can compute
partition functions and marginals at cost
$O(n \cdot 2^{\mathrm{tw}})$~\cite{KollerFriedman:2009}.
Computing $P(C(x)>T)$ requires additional cost-dependent bookkeeping,
but remains classically tractable for this $n=20$ validation instance.
The quantum pipeline therefore
does not outperform the best exact classical method on this specially
chosen validation problem.  Its demonstrated advantage is instead over
classical Monte Carlo estimation, which is the relevant operational
baseline when exact inference is unavailable, too costly, or not trusted
for large correlated supply-chain models.

The same bounded-treewidth structure is also what makes the present instance
valuable as a benchmark.  Because exact probabilities are available, the
full quantum pipeline can be validated against ground truth rather than
only compared empirically against other stochastic estimators.  In this
sense, the instance is deliberately conservative: it allows a clean
diagnosis of the state-preparation circuit, the oracle construction,
and the amplitude-estimation procedure before moving to regimes where
exact classical verification is no longer feasible.

A second qualification concerns the distinction between statistical
and systematic error.  \ac{mlae} reduces sampling error, but it cannot
remove the two deterministic floors defined in
Sec.~\ref{sec:error_decomposition}: the state-preparation bias and the
structural under-marking error $\epsilon_\mathrm{miss}$
(Eq.~\eqref{eq:eps_miss}).  For the present model the under-marking
residual is small relative to the total tail probability at
$T_\mathrm{struct}$, but it remains a fixed oracle-induced error.

The model-oracle path isolates the oracle contribution at
$T_\mathrm{ref}$. Because it marks states according to the complete
classical cost table, it introduces no structural under-marking while
retaining the same state-preparation circuit. Its agreement with
$Q^*(C>T_\mathrm{ref})$ confirms that the model oracle contributes no
additional error. For the structural oracle, the gap between its amplitude
and $Q^*(C>T_\mathrm{struct})$ directly measures the under-marking
contribution. Reporting these deterministic gaps alongside the statistical
estimation error is essential for a faithful empirical quantum-acceleration
claim. The relevant question is not only how fast the estimator converges,
but also to what quantity it converges.

\subsection{Scaling Prospects and the Role of Graph Complexity}

Scaling is another central consideration when assessing the utility of quantum algorithms.
The central scaling issue here is not the number of variables alone, but the
graph complexity of the underlying supply-chain model.  For sparse
graphs with low treewidth, exact classical inference remains efficient.
For loopy graphs with dense local structure, the cost of variable
elimination grows as $O(n2^{\mathrm{tw}})$ and becomes prohibitive once
the effective treewidth is moderately large.  In supply-chain models,
this transition is expected when cross-tier dependencies, regional
shock factors, and shared suppliers introduce multiple overlapping
cycles and dense clusters.  In that regime, exact inference ceases to
be the relevant classical competitor, and Monte Carlo methods become
the practical baseline.

Monte Carlo estimation also becomes more difficult as the event becomes
rarer.  For a tail probability $p=P(C>T)$, the sampling cost needed to
resolve the event scales unfavourably with $1/p$, while the convergence
rate remains $O(N^{-1/2})$. Variance-reduction methods can improve constants
but generally retain the square-root sampling rate. This is precisely the
regime in which amplitude estimation is attractive. For the fixed schedule
used here, however, \ac{mlae}'s RMSE also scales as
$\Theta(N_\mathrm{query}^{-1/2})$; the ideal
$O(N_\mathrm{query}^{-1})$ rate requires the maximum depth to grow with the
query budget (Sec.~\ref{sec:results_schedule_scaling}).

The two constructive components introduced in this work scale
polynomially in the sparse-graph regime.  The \ac{bpacl}
state-preparation circuit uses $O(n)$ conditional rotations after
classical belief propagation and Chow-Liu projection.  For sparse Ising
graphs with $|E|=O(n)$, the preprocessing cost scales as
$O(|E| \times \mathrm{iterations})$, which is linear in $n$ for a fixed
number of message-passing iterations.  The structural oracle requires
$n+O(\log |G|)$ qubits and has depth $O(n\log |G|)$, where $|G|$ is the
size of the largest rule group.  Thus, provided the rule structure
contains a bounded number of threshold rules with sub-linear group
size, both state preparation and oracle evaluation remain polynomial as
the model grows. The accuracy of this scaling picture depends on
sparsity. \ac{bp} accuracy is controlled primarily by graph density,
coupling strength, and cycle structure rather than by $n$ itself.

There are nevertheless important limitations.  If the graph
becomes dense or contains many short, strongly coupled cycles, \ac{bp}
may no longer provide sufficiently accurate marginals.  Within the
treewidth-only family tested in Sec.~\ref{sec:state_prep_robustness},
this degradation already becomes appreciable at treewidth 3 to 5, well
short of a dense graph in absolute terms.  The same section also
varies coupling strength directly at fixed topology and finds that
$D_{\mathrm{KL}}$ can change by one to two orders of magnitude within
the limited set of cases tested (Fig.~\ref{fig:graph_family}). The
resulting state-preparation bias would enlarge the deterministic floor below
which \ac{mlae} cannot improve.  Dense models would require more
expressive classical inference methods, such as cluster belief
propagation, junction-tree inference on reduced-width subgraphs, tensor
network methods, or alternative quantum state-preparation architectures.

This creates a narrow, and currently uncharacterised, window for
practical benefit. Exact graphical-model inference remains efficient at
fixed treewidth, although exact tail-probability evaluation also depends
on the cost representation. A useful quantum speedup would therefore
require graph and cost structures for which exact tail inference becomes
genuinely expensive, beyond the regime studied here.

A further limitation is that the present results are obtained in noiseless
simulation. On real hardware, errors accumulate with the depth of the
Chow-Liu circuit, the structural oracle, and the repeated Grover operators
used by \ac{mlae}. At the $n=50$--$100$ scale, this depth is beyond the
capability of present \ac{nisq} devices. Indeed, even on the present $n=20$
benchmark, the $k=3$ Grover circuit requires $2\,323$ CX gates after basis
decomposition, placing it in the fault-tolerant regime under
current two-qubit gate error rates.
Realising the asymptotic query-complexity speedup will therefore require
fault-tolerant \ac{qpu}s or substantially lower-depth constructions,
together with an analysis of the regime in which the speedup survives noise.

\section{Conclusions}
\label{sec:conclusions}

This paper developed and tested a quantum amplitude-estimation pipeline for
correlated combinatorial supply-chain risk models. Its two constructive
components address central implementation barriers: the \ac{bpacl} method
prepares a correlated Ising distribution without Monte Carlo sampling, while
the structural oracle encodes threshold-rule costs directly as a reversible
Boolean circuit. The resulting state-preparation circuit has linear size in
the sparse-graph regime, and its accuracy remains high across the graph
topologies tested. Together, these components provide a pipeline without
hidden state-preparation sampling, whose state-preparation and oracle errors
can be diagnosed separately.

On the $n=20$ validation benchmark, \ac{mlae} achieved lower statistical error
than four classical Monte Carlo baselines at medium-to-large budgets and at
two levels of event rarity. Under the fixed Grover-depth schedule used here,
however, this improvement is a constant factor rather than the ideal
amplitude-estimation rate: any fixed schedule retains
$\operatorname{RMSE}=\Theta(N_\mathrm{query}^{-1/2})$, with greater depth
changing only the prefactor. At the rarer structural threshold, increasing
the budget eventually exposes a deterministic accuracy floor produced by the
Chow--Liu state-preparation bias and structural oracle under-marking. Separating
these effects from sampling noise is therefore essential when assessing
estimator convergence.

Because the benchmark permits exact enumeration, the complete pipeline can be
validated against ground truth. At matched budgets, it achieves lower
estimation error than the four classical sampling-based estimators considered
here. The relevant regime for potential quantum utility begins when graph
complexity makes exact inference impractical and sampling becomes the classical
baseline. Two limitations define the
immediate scope of the proposed pipeline. The \ac{bpacl} approximation may
need to be strengthened or replaced for dense, strongly loopy graphs, while
the structural oracle currently covers only threshold-rule costs. Moreover,
the reported gate counts place multi-iteration \ac{mlae} in the fault-tolerant
rather than near-term regime. Future work should therefore prioritise more
expressive state-preparation methods and lower-depth oracle constructions;
realising the ideal query scaling will additionally require an \ac{mlae}
schedule whose maximum depth grows with the query budget. The present results
establish the statistical benefit and deterministic limitations of the
proposed pipeline, while identifying the circuit-depth and scaling
requirements that must be met before a practical quantum speedup in query
complexity can be established.

\data{Code and data are available upon request to the authors.}

\appendix
\section{Belief Propagation on the Ising Factor Graph}
\label{app:bp}

Equation~\eqref{eq:bp_update} is the sum-product update rule applied to
the pairwise Markov random field in~\eqref{eq:ising_factorisation}
\cite{Yedidia:2003,Aji:2000}. Here we write this update in the log
domain, as implemented in the code, and derive the pairwise belief used
to compute $p_{ij}$.

\subsection*{Log-domain message update}

To evaluate Eq.~\eqref{eq:bp_update} numerically, we work in
log-domain: products of probabilities are replaced by sums of
log-probabilities, which avoids numerical underflow when many messages
are multiplied together~\cite{KollerFriedman:2009}.
The key term in the update rule is the product
$\prod_{k\in\partial i\setminus j}\mu_{k\to i}(x_i)$, which
aggregates all incoming messages at node $i$ \emph{excluding} the
message from $j$.
Excluding $j$ is essential: including it would feed $j$'s own
information back through $i$, creating circular reasoning that
corrupts the marginals.

In log-domain, this product over $\partial i\setminus j$ becomes a
sum of log-messages.
Absorbing the unary factor $\phi_i(x_i)=e^{h_i x_i}$ (which
contributes $h_i$ only when $x_i=1$), we define the
\emph{cavity log-weights} for the directed message $i\!\to\!j$:
\begin{align}
  \ell_0^{(i\setminus j)}
  &= \sum_{k\in\partial i\setminus j}\log\mu_{k\to i}(0),
  \label{eq:cav0}\\
  \ell_1^{(i\setminus j)}
  &= h_i + \sum_{k\in\partial i\setminus j}\log\mu_{k\to i}(1).
  \label{eq:cav1}
\end{align}

Expanding the sum in~\eqref{eq:bp_update} over $x_i\in\{0,1\}$ and
using $\psi_{ij}(x_i,x_j)=\exp(J_{ij}x_ix_j)$ — which equals $1$
whenever $x_ix_j=0$ — gives
\begin{equation}
\begin{aligned}
  \mu_{i\to j}(0)
  &\propto
  \exp(\ell_0^{(i\setminus j)})
  + \exp(\ell_1^{(i\setminus j)}), \\
  \mu_{i\to j}(1)
  &\propto
  \exp(\ell_0^{(i\setminus j)})
  + \exp(\ell_1^{(i\setminus j)}+J_{ij}).
\end{aligned}
\label{eq:msg_expand}
\end{equation}
Equation~\eqref{eq:msg_expand} is the update rule applied to every directed 
message at each iteration. All messages are recomputed simultaneously from 
their current values (synchronous schedule) and the procedure repeats until 
the maximum message change falls below a tolerance.
These are evaluated using log-sum-exp, then normalised to
\[
  \mu_{i\to j}(0)+\mu_{i\to j}(1)=1.
\]
Writing $\ell_s=\ell_s^{(i\setminus j)}$ for brevity,
\begin{equation}
  \log\tilde\mu_{i\to j}(s)
  =
  \log\!\left[
    \exp(\ell_0)
    +
    \exp\!\left(\ell_1+J_{ij}[s=1]\right)
  \right],
  \label{eq:log_msg_raw}
\end{equation}
and $\tilde\mu_{i\to j}$ is normalised to obtain $\mu_{i\to j}$.

After convergence, the one-body marginal in~\eqref{eq:bp_marginal} follows 
from the node belief
\[
  b_i(x_i)
  \propto
  \phi_i(x_i)\prod_{k\in\partial i}\mu_{k\to i}(x_i).
\]
Substituting $\phi_i(x_i)=e^{h_i x_i}$ and evaluating at each value
of $x_i\in\{0,1\}$ gives
\begin{align*}
  b_i(0) &\propto \prod_{k\in\partial i}\mu_{k\to i}(0), \\
  b_i(1) &\propto e^{h_i}\!\prod_{k\in\partial i}\mu_{k\to i}(1).
\end{align*}
Writing $L_s = \log b_i(s)$ and using
\[
  p_i = e^{L_1}/(e^{L_0}+e^{L_1}) = \sigma(L_1-L_0), 
\]
where $\sigma(z)$ denotes the logistic function, one finds the log-odds
difference,
\[
  L_1 - L_0
  = h_i + \sum_{k\in\partial i}\log\frac{\mu_{k\to i}(1)}{\mu_{k\to i}(0)},
\]
which is precisely the argument of $\sigma$ in Eq.~\eqref{eq:bp_marginal}.

\subsection*{Pairwise belief on Ising edges}

For each edge $(i,j)\in E$ we want $p_{ij}=P(x_i=1,x_j=1)$.
By analogy with the node belief, the joint belief on the pair
$(x_i,x_j)$ combines the cavity contributions from both endpoints
with the edge factor $\psi_{ij}$:
\begin{equation}
\begin{split}
  b_{ij}(x_i,x_j)
  &\propto
  \phi_i(x_i)\,\phi_j(x_j)\,\psi_{ij}(x_i,x_j)\\
  &\quad\times
  \!\!\prod_{k\in\partial i\setminus j}\!\!\mu_{k\to i}(x_i)\;
  \!\!\prod_{k\in\partial j\setminus i}\!\!\mu_{k\to j}(x_j).
\end{split}
\label{eq:pair_belief}
\end{equation}
Each product excludes the mutual edge $(i,j)$: including it on
both sides would count the same interaction twice.
On a tree, $b_{ij}(x_i,x_j)\propto P(x_i,x_j)$ at convergence, so
normalising over all four states recovers the true joint marginal.
On a loopy graph, convergence to the correct fixed point is not
guaranteed, and $b_{ij}$ is generally only an approximation to the
joint marginal rather than $P(x_i,x_j)$ itself, as discussed below.

Taking the log of~\eqref{eq:pair_belief} and absorbing $\phi_i$,
$\phi_j$ into the cavity log-weights — defined symmetrically for
node $j$ by exchanging $i\leftrightarrow j$ in
Eqs.~\eqref{eq:cav0}--\eqref{eq:cav1} — gives the unnormalised
pairwise log-belief:
\begin{equation}
  \log b_{ij}(x_i,x_j)
  =
  \ell_{x_i}^{(i\setminus j)}
  +
  \ell_{x_j}^{(j\setminus i)}
  +
  J_{ij}x_ix_j,
  \quad (x_i,x_j)\in\{0,1\}^2.
  \label{eq:joint_belief}
\end{equation}
Reading off the $(x_i,x_j)=(1,1)$ entry and normalising over all
four states gives
\begin{equation}
  p_{ij}
  =
  \frac{
    \exp\!\left(
      \ell_1^{(i\setminus j)}
      +
      \ell_1^{(j\setminus i)}
      +
      J_{ij}
    \right)
  }{
    \displaystyle
    \sum_{a,b\in\{0,1\}}
    \exp\!\left(
      \ell_a^{(i\setminus j)}
      +
      \ell_b^{(j\setminus i)}
      +
      J_{ij}ab
    \right)
  }.
  \label{eq:p11_formula}
\end{equation}

For non-edges $(i,j)\notin E$, no pairwise cavity computation is used.
Instead, we approximate
\[
  p_{ij}\approx p_i p_j.
\]
This is appropriate when the mutual information between $i$ and $j$ is
negligible; for all dropped edges in \ac{scm20}, it is below
$2\times10^{-4}$\,nats, as discussed in Sec.~\ref{sec:state_prep}.

\subsection*{Exactness and convergence}

On a tree, removing an edge $(i,j)$ separates the graph into two
independent subtrees. The message $\mu_{i\to j}$ is then the exact
marginal contribution of the subtree on the $i$ side, with no circular
flow of information. Therefore, \ac{bp} returns exact marginals after one
forward-backward pass \cite{KollerFriedman:2009}.

In loopy graphs, messages may circulate around cycles, so convergence to
the correct fixed point is not guaranteed in general \cite{Murphy:1999}.
When the independent cycles are few and short --- as in \ac{scm20},
where the coupling graph has cyclomatic number $L=4$ confined to small,
low-treewidth blocks --- this feedback remains limited. In sparse
graphs with moderate couplings, the resulting fixed point is often
empirically accurate \cite{Yedidia:2003}.

\section{Correctness of the Chow-Liu State-Preparation Circuit}
\label{app:chow_liu}

We prove that the circuit $\mathcal{R}$ defined by
Eqs.~\eqref{eq:root_ry} and~\eqref{eq:cond_ry} prepares the state
\begin{equation}
  \mathcal{R}\,|0\rangle^{\otimes n}
  =
  \sum_{x\in\{0,1\}^n}
  \sqrt{Q^*(x)}\,|x\rangle,
  \label{eq:R_claim}
\end{equation}
where $Q^*(x)$ is the Chow-Liu factorisation~\eqref{eq:chow_liu_factorisation}.
The argument is an induction on the directed tree ordering.

We first verify that the block in Eq.~\eqref{eq:cond_ry} acts as a
controlled-$R_y$ on qubit $i$.
When the parent qubit is in basis state $|s\rangle$
with $s\in\{0,1\}$, the sequence $\mathrm{CX}-R_y(\beta_i)-\mathrm{CX}-R_y(\alpha_i)$
applies to qubit $i$ as follows.
For $s=0$, both $\mathrm{CX}$ gates are inactive, so the operations result in a net rotation
$R_y(\beta_i)R_y(\alpha_i) = R_y(\alpha_i+\beta_i) = R_y(\theta_{i|0})$.
For $s=1$ each $\mathrm{CX}$ sandwiches $R_y(\beta_i)$ with a bit-flip,
and using the identity $X R_y(\theta) X = R_y(-\theta)$ gives
$R_y(\alpha_i-\beta_i) = R_y(\theta_{i|1})$.
In both cases, starting from $|0\rangle_i$,
\begin{equation}
  |s\rangle_{\pi(i)}\,|0\rangle_i
  \;\longmapsto\;
  |s\rangle_{\pi(i)}
  \otimes
  \sum_{x_i}\sqrt{P(x_i\mid x_{\pi(i)}=s)}\,|x_i\rangle,
  \label{eq:block_action}
\end{equation}
where the sum over $x_i\in\{0,1\}$ follows from $R_y(\theta_{i|s})|0\rangle
= \sqrt{1-p_s}\,|0\rangle + \sqrt{p_s}\,|1\rangle$
with $p_s = P(x_i=1\mid x_{\pi(i)}=s)$.

\subsection*{Inductive proof}
Let the nodes be processed in any ordering consistent with the directed
Chow-Liu tree (parent before children).
After processing nodes $v_1=r,v_2,\ldots,v_k$, we claim the circuit state is
\begin{equation}
  \sum_{x_{v_1},\ldots,x_{v_k}}
  \sqrt{Q_k^*(x)}\;
  |x_{v_1},\ldots,x_{v_k}\rangle
  \otimes
  |0\rangle^{\otimes(n-k)},
  \label{eq:inductive_state}
\end{equation}
where $Q_k^*(x) = P(x_r)\prod_{j=2}^{k}P(x_{v_j}\mid x_{\pi(v_j)})$.

\textit{Base case} ($k=1$, root only).
The root gate gives
\begin{equation}
  R_y(\theta_r)\,|0\rangle
  = \sqrt{1-p_r}\,|0\rangle + \sqrt{p_r}\,|1\rangle
  = \sum_{x_r}\sqrt{P(x_r)}\,|x_r\rangle,
\end{equation}
which matches~\eqref{eq:inductive_state} for $k=1$.

Suppose~\eqref{eq:inductive_state} holds after $k$ nodes.
Node $v_{k+1}$ has its parent $\pi(v_{k+1})$ already processed,
so the parent qubit is in a superposition of basis states.
By linearity, applying~\eqref{eq:block_action} to the tensor product
extends the sum by one factor:
\begin{equation}
  Q_{k+1}^*(x)
  = Q_k^*(x)\cdot P\!\bigl(x_{v_{k+1}}\mid x_{\pi(v_{k+1})}\bigr).
\end{equation}
The state after $n$ nodes therefore has
$Q_n^*(x) = Q^*(x)$, which establishes~\eqref{eq:R_claim}.\hfill$\square$

\section{Structural Oracle: Gate-Level Circuit Construction}
\label{app:oracle_circuit}

This appendix describes the reversible circuit used to implement the
structural oracle $\mathcal{F}_T^\mathrm{struct}$ given by Eq.~\ref{eq:struct_oracle_action} and 
introduced in Sec.~\ref{sec:oracle_struct}. The oracle receives a supply-chain state
$\ket{x}$ and writes one output flag qubit. The flag is set to one if
at least one of the two structural failure rules is satisfied.
For a system like \ac{scm20}, $q_{20}$ corresponds to the output flag, 
while the remaining seven ancilla qubits are scratch qubits. 
A key requirement is that all scratch qubits
must be returned to $\ket{0}$ at the end of the oracle.

Two rules are implemented in the circuit:

\begin{itemize}[topsep=4pt, itemsep=2pt, parsep=0pt]
\item $\mathrm{rule}_1$ checks whether at least three Tier-1 suppliers
fail. This is implemented using a ripple-carry counter followed
by a comparator.
\item $\mathrm{rule}_2$ checks whether at least one manufacturing plant
and at least one port fail. This is implemented using two OR
gadgets followed by a Toffoli gate.
\end{itemize}

Because these computations act on quantum superpositions, they cannot
leave intermediate information behind in the scratch register. Any such
leftover information would entangle the scratch qubits with the data
register and would therefore corrupt the coherence of the state
preparation unitary $\mathcal{R}$ from Sec.~\ref{sec:state_prep}.
We therefore use the standard compute--copy--uncompute construction
of Bennett~\cite{Bennett:1973}. First, the rule values are computed
into scratch qubits. Then their logical OR is copied into the output
flag. Finally, the rule computations are undone gate by gate.

To add more clarity, we also include here the qubit layout. 
The SCM20 instance uses $20$ data qubits and $8$ ancilla qubits, giving
a total of $28$ qubits. The allocation is shown in
Table~\ref{tab:qubit_layout}.

\subsection*{Rule 1: threshold failure among Tier-1 suppliers}
The first rule checks whether at least three of the five Tier-1 supplier
nodes fail. For SCM20, the relevant group is
\begin{equation*}
G_1 = \{q_0,q_1,q_2,q_3,q_4\},
\end{equation*}
corresponding to supplier nodes 1--5. The threshold is
\begin{equation*}
k = 3.
\end{equation*}
The circuit first counts the number of failed nodes in $G_1$. Since the
group has five elements, the counter must represent values from $0$ to
$5$. Three counter qubits are therefore sufficient:
\begin{equation*}
(c_0,c_1,c_2) = (q_{21},q_{22},q_{23}),
\end{equation*}
with $c_0$ the least significant bit.

For each input qubit $x_i \in G_1$, the circuit performs a controlled
increment of the counter. The increment is applied only when $x_i = 1$.
The counter update is carried out from most significant bit to least
significant bit, so that each carry condition uses the original lower
counter bits.

For a three-bit counter, one controlled increment consists of
\begin{equation*}
\begin{aligned}
c_2 &\leftarrow c_2 \oplus (x_i \wedge c_1 \wedge c_0), \\
c_1 &\leftarrow c_1 \oplus (x_i \wedge c_0), \\
c_0 &\leftarrow c_0 \oplus x_i .
\end{aligned}
\end{equation*}
These three updates are implemented in order (MSB first) by
\begin{equation*}
  \mathrm{C^3X}(x_i,\,c_0,\,c_1 \to c_2),\quad
  \mathrm{CCX}(x_i,\,c_0 \to c_1),\quad
  \mathrm{CX}(x_i \to c_0).
  \label{eq:increment_gates}
\end{equation*}
Each increment therefore uses one 3-controlled-$X$, one Toffoli, and one CNOT gate.

After all five increments, the counter stores
\begin{equation*}
v = \sum_{i \in G_1} x_i .
\end{equation*}
The comparator then writes the threshold result into $q_{24}$:
\begin{equation*}
q_{24} =
\mathbf{1}[v \geq 3].
\end{equation*}

For a three-bit counter with possible values $v \in \{0,1,2,3,4,5\}$,
the threshold condition can be written as
\begin{equation}
v \geq 3
\Longleftrightarrow
c_2 \vee (c_1 \wedge c_0).
\label{eq:comparator_k3}
\end{equation}
The two cases on the right-hand side are mutually exclusive in this
range: $c_1 c_0 = 1$ detects $v=3$, while $c_2=1$ detects $v=4$ or
$v=5$. Hence the OR can be implemented by two XOR-style reversible
updates:
\begin{equation*}
\mathrm{CX}(c_2 \to q_{24}),
\qquad
\mathrm{CCX}(c_1,c_0 \to q_{24}).
\end{equation*}

The comparator is self-inverse, since applying the same XOR updates
twice restores $q_{24}$.

\subsection*{Rule 2: simultaneous plant and port failure}
The second rule checks whether there is at least one failed plant and
at least one failed port.

The plant group is
\begin{equation*}
G_{2A} = \{q_{10},q_{11},q_{12}\},
\end{equation*}
corresponding to nodes 11--13. The port group is
\begin{equation*}
G_{2B} = \{q_{13},q_{14}\},
\end{equation*}
corresponding to nodes 14--15.

The circuit computes
\begin{equation*}
a_A = \bigvee_{q_i \in G_{2A}} q_i
\qquad \text{into } q_{25},
\end{equation*}
and
\begin{equation*}
a_B = \bigvee_{q_i \in G_{2B}} q_i
\qquad \text{into } q_{26}.
\end{equation*}
It then combines these two intermediate values with a Toffoli gate:
\begin{equation*}
q_{27} \leftarrow q_{27} \oplus (q_{25} \wedge q_{26}).
\end{equation*}
Thus $q_{27}$ stores the value of $\mathrm{rule}_2$.

The OR operation itself is implemented reversibly using De~Morgan's law.
For inputs $x_1,\ldots,x_m$ and a scratch qubit $a$ initially in
$\ket{0}$, the OR gadget performs the following steps:

\begin{enumerate}[topsep=4pt, itemsep=2pt, parsep=0pt, label=\arabic*.]
\item Apply $X$ gates to all input qubits.
\item Apply a multi-controlled $X$ gate from the input qubits to $a$.
\item Apply $X$ gates again to restore the input qubits.
\item Apply $X$ to $a$.
\end{enumerate}

This computes
\begin{equation*}
a = x_1 \vee \cdots \vee x_m.
\end{equation*}
The inverse OR gadget is obtained by applying the same four steps in
reverse order.

For SCM20, the plant OR uses a three-controlled gate and the port OR
uses a two-controlled gate. The two OR gadgets are later uncomputed, so
their scratch qubits $q_{25}$ and $q_{26}$ return to $\ket{0}$.

\begin{table}[t]
\centering
\begin{tabular}{llp{6.5cm}}
\toprule
Qubits & Count & Role \\
\midrule
$q_0$--$q_{19}$    & 20 & Data register, corresponding to supply-chain nodes 1--20 \\
$q_{20}$           & 1  & Output flag used by QAE \\
$q_{21}$--$q_{23}$ & 3  & Counter register for $\mathrm{rule}_1$ \\
$q_{24}$           & 1  & Result qubit for $\mathrm{rule}_1$ \\
$q_{25}$           & 1  & OR result for plant group $G_{2A}$ \\
$q_{26}$           & 1  & OR result for port group $G_{2B}$ \\
$q_{27}$           & 1  & Result qubit for $\mathrm{rule}_2$ \\
\bottomrule
\end{tabular}
\caption{Qubit allocation for the SCM20 structural oracle.}
\label{tab:qubit_layout}
\end{table}

\subsection*{Combining the two rule flags}

At this stage, the circuit has computed
\begin{equation*}
q_{24} = \mathbf{1}[\mathrm{rule}_1(x)]
\qquad \text{and} \qquad
q_{27} = \mathbf{1}[\mathrm{rule}_2(x)].
\end{equation*}
The oracle output flag $q_{20}$ must be set if either rule fires:
\begin{equation*}
q_{20}
=
q_{24} \vee q_{27}.
\end{equation*}

Since $q_{20}$ starts in $\ket{0}$, this OR is written reversibly using
\begin{equation*}
a \vee b = (a \oplus b) \oplus (a \wedge b).
\end{equation*}
The corresponding gates are
\begin{equation*}
\mathrm{CX}(q_{24} \to q_{20}), \qquad
\mathrm{CX}(q_{27} \to q_{20}), \qquad
\mathrm{CCX}(q_{24},q_{27} \to q_{20}).
\end{equation*}

After these three gates, the output qubit $q_{20}$ contains the final
oracle value. The remaining scratch qubits still contain intermediate
information and must therefore be uncomputed.

\subsection*{Compute--copy--uncompute structure}
The full oracle is organised into three phases.

\emph{Phase 1.} Compute the rule values:
First, the circuit computes $\mathrm{rule}_1$ into $q_{24}$. This is
done by applying five controlled increments to the counter
$q_{21}$--$q_{23}$, followed by the two comparator gates from
Eq.~\eqref{eq:comparator_k3}. Second, the circuit computes $\mathrm{rule}_2$ into $q_{27}$. This is
done by computing the plant OR into $q_{25}$, computing the port OR into
$q_{26}$, and applying a Toffoli gate from $q_{25}$ and $q_{26}$ to
$q_{27}$.

\emph{Phase 2.} Copy the final answer to the output flag:
The two rule flags $q_{24}$ and $q_{27}$ are combined into the output
qubit $q_{20}$ using two CNOT gates and one Toffoli gate.

\emph{Phase 3.} Uncompute all scratch information:
Finally, the circuit applies the inverse of Phase 1 in reverse order.
It first uncomputes $\mathrm{rule}_2$ by undoing the Toffoli gate and
then undoing both OR gadgets. It then uncomputes $\mathrm{rule}_1$ by
undoing the comparator and applying the five inverse controlled
increments. At the end of this phase, all scratch qubits have returned to
$\ket{0}$ and only the output flag $q_{20}$ remains changed.

For the SCM20 instance, the structural oracle contains $67$ gates and
has depth $36$.
The oracle is composed exclusively of $X$, CNOT, Toffoli, and
3-controlled-$X$ gates, consistent with the gate set stated in
Sec.~\ref{sec:oracle_struct}, and is directly executable on a \ac{qpu}
without classical enumeration of supply-chain failure states.

\begin{figure*}[t]
  \centering
  \includegraphics[width=\textwidth]{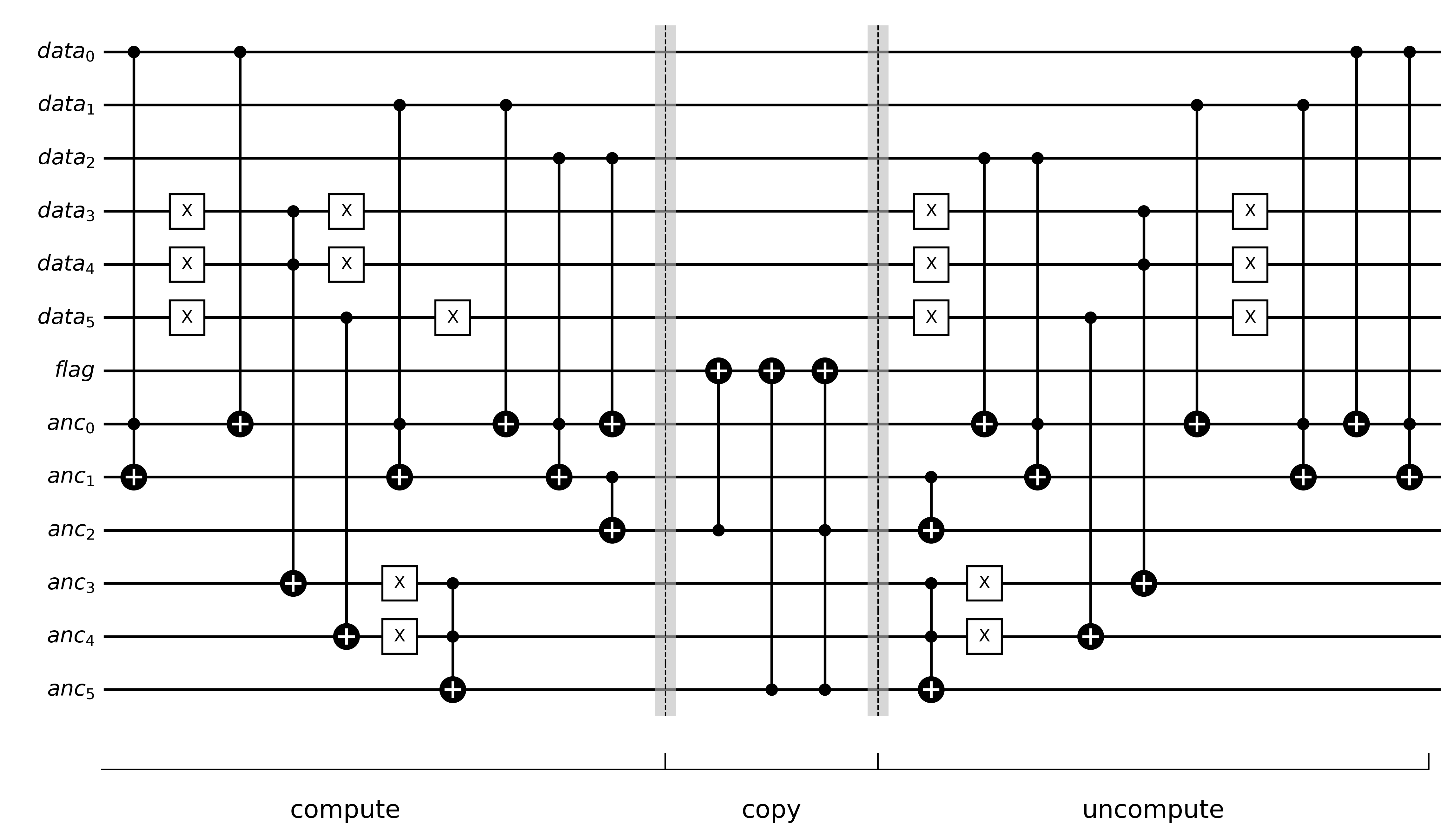}
  \caption{\label{fig:oracle_toy}%
    Structural oracle circuit for a small example instance. The
    circuit is generated with the same function used to build the
    SCM20 oracle but with smaller rule groups so every gate is
    legible. The three phases --- compute, copy, and uncompute --- are marked
    below the circuit. The 3-controlled-$X$ gate used in the SCM20
    circuit does not appear here because the smaller counter does not
    require one.}
\end{figure*}

To make this structure concrete we show in Fig.~\ref{fig:oracle_toy}
the circuit for a small example instance. The circuit is generated
with the same function used to build the SCM20 oracle but with smaller
rule groups so every gate remains visible: a threshold group of three
nodes with $k=2$, and an and-any rule over groups of two and one
nodes. The three phases --- compute,
copy, and uncompute --- are marked below the circuit and the symmetry
between the compute phase and the uncompute phase is directly visible.

\subsection*{Resource scaling with Grover depth}

The combined $A = F_T \cdot (R \otimes I)$ circuit with the structural
oracle has substantial resource requirements, including
52~CX, 17~CCX (Toffoli), and 12~MCX gates. After transpilation to CNOT
and single-qubit gates with full compiler optimisation, the full
$\mathcal{A}$ circuit requires 283~CX gates. Each additional Grover
iteration appends two copies of $\mathcal{A}$ and one $S_0$ reflection,
so gate counts grow additively with $k$ (Table~\ref{tab:gate_resources}),
reaching 963~CX at $k=1$ and 2\,323~CX at $k=3$.
Table~\ref{tab:gate_resources} reports the full per-depth breakdown,
including the CCX and multi-controlled-$X$ counts and circuit depth,
which grow at the same $(2k{+}1)$ rate as the CX count since every
quantity scales with the number of $\mathcal{A}$/$\mathcal{A}^\dagger$
copies in $\mathcal{Q}^k\mathcal{A}$. These values are reported only for
the structural oracle, since the model oracle's DiagonalGate
construction is not intended as an executable circuit
(Sec.~\ref{sec:oracle_design}).

\begin{table}[h]
\centering
\caption{\label{tab:gate_resources}%
  Circuit resources for $\mathcal{Q}^k\mathcal{A}$ (structural oracle) as
  a function of Grover depth $k$, transpiled to the $\{\mathrm{CX}, U\}$
  basis at optimisation level 3.  CX and depth are post-transpile;
  CCX/MCX are raw pre-transpile counts on the untranspiled circuit,
  reported to show how many multi-controlled gates the construction
  contains at each depth.  Values follow the additive relation
  $N(\mathcal{Q}^k\mathcal{A}) \approx (2k{+}1)N(\mathcal{A}) +
  k\,N(S_0)$; see Sec.~\ref{sec:budget_gates}.}
\begin{tabular}{ccccc}
\toprule
$k$ & CX & CCX & MCX & Depth \\
\midrule
0 & 283   & 17  & 12 & 356   \\
1 & 963   & 51  & 36 & 1\,176 \\
2 & 1\,643 & 85  & 60 & 1\,996 \\
3 & 2\,323 & 119 & 84 & 2\,816 \\
\bottomrule
\end{tabular}
\end{table}

\section{Fisher-Information Derivation for Fixed-Depth MLAE Schedules}
\label{app:schedule_scaling}

This appendix derives the fixed-schedule RMSE rate quoted in
Sec.~\ref{sec:results_schedule_scaling}, Eq.~\eqref{eq:rmse_fixed_schedule}.

For one shot at Grover depth $k$, the Fisher information for $\theta_a$ is
\begin{equation}
  I_k(\theta_a)
  = \frac{\bigl(\mathrm{d}p_k/\mathrm{d}\theta_a\bigr)^2}{p_k(1-p_k)}
  = 4(2k+1)^2,
  \label{eq:fisher_k}
\end{equation}
independent of $\theta_a$, obtained directly from $p_k$ (Eq.~\eqref{eq:pk}).
For a fixed schedule $\mathcal{K}$ with $N_k =
N_\mathrm{shots}/|\mathcal{K}|$ shots allocated equally to each depth,
the total Fisher information is
\begin{equation}
  I(\theta_a)
  = \sum_{k\in\mathcal{K}} N_k \cdot 4(2k+1)^2
  = \frac{4N_\mathrm{shots}}{|\mathcal{K}|}\sum_{k\in\mathcal{K}}(2k+1)^2,
  \label{eq:fisher_total}
\end{equation}
which is \emph{linear} in $N_\mathrm{shots}$ for any fixed $\mathcal{K}$,
exactly as $N_\mathrm{query}$ is (Eq.~\eqref{eq:n_query}).  Since both
quantities are linear in $N_\mathrm{shots}$ for any fixed schedule, one
is proportional to the other,
\begin{equation}
  I(\theta_a) \;\propto\; N_\mathrm{query}.
  \label{eq:fisher_prop_query}
\end{equation}
The Cram\'er--Rao bound gives
\begin{equation}
  \mathrm{Var}(\hat\theta_a)
  \geq \frac{1}{I(\theta_a)}
  = \Omega\!\left(\frac{1}{N_\mathrm{query}}\right),
\end{equation}
for unbiased estimators, ruling out variance that decreases faster than
$N_\mathrm{query}^{-1}$. Under the standard regularity and local
identifiability conditions for maximum-likelihood estimation, the
\ac{mlae} estimator is asymptotically efficient and attains this scaling:
\begin{equation}
  \mathrm{Var}(\hat\theta_a)
  = \Theta\!\left(\frac{1}{N_\mathrm{query}}\right).
  \label{eq:var_query}
\end{equation}
Taking the square root, and using the asymptotic consistency of the
maximum-likelihood estimator, gives the RMSE rate quoted in the main text,
Eq.~\eqref{eq:rmse_fixed_schedule}.

Nothing in this chain of reasoning depends on how deep the schedule is.
The result holds regardless of how large $k_\mathrm{max}$ is, as long
as the set of depths used stays fixed while the shot budget grows.  A
deeper fixed schedule changes only the constant prefactor in
Eq.~\eqref{eq:fisher_total}, not the exponent in
Eq.~\eqref{eq:rmse_fixed_schedule}.  The same reasoning also does not
depend on the equal-shots-per-depth allocation used in the main text.
Any fixed proportional split $N_k = w_k N_\mathrm{shots}$ leaves both
$I(\theta_a)$ and $N_\mathrm{query}$ linear in $N_\mathrm{shots}$, so
Eq.~\eqref{eq:fisher_prop_query} and the resulting RMSE scaling in
Eq.~\eqref{eq:rmse_fixed_schedule} remain unchanged for any fixed allocation
across $\mathcal{K}$.

\bibliographystyle{unsrt}
\bibliography{bibliography}

\end{document}